\documentclass[lineno]{jfm}

\usepackage{graphicx}
\usepackage{amsmath}
\usepackage{amssymb}
\usepackage{pifont}
\usepackage{txfonts}
\usepackage{psfrag}
\usepackage[round]{natbib}
\usepackage[T1]{fontenc}
\usepackage{lmodern}
\usepackage{tikz}
\usetikzlibrary{shapes.geometric,shapes.symbols}
\usepackage{scalerel}
\usepackage{upgreek}
\usepackage{booktabs}
\usepackage{verbatim}
\usepackage{framed}
\usepackage{fancyvrb}
\usepackage{varwidth}
\usepackage{bbding}
\usepackage{subfigure}
\usepackage{multirow}
\usepackage[normalem]{ulem}
\usepackage{multicol}
\usepackage{xcolor}
\usepackage{breqn}
\usepackage{soul}
\usepackage{dirtytalk}
\usepackage{upquote}
\usepackage{authblk}
\usepackage{array}
\newcolumntype{P}[1]{>{\centering\arraybackslash}p{#1}}
\definecolor{violet}{rgb}{0.54, 0.17, 0.89}
\definecolor{golden}{rgb}{0.99, 0.76, 0.0}

\usepackage{hyperref}
\hypersetup{
    colorlinks = true,
    urlcolor   = blue,
    citecolor  = black,
}

\newcommand{\RomanNumeralCaps}[1]
\linenumbers

\newcommand{\rahul}[1]{{\color{black}#1}}
\newcommand{\rahuld}[1]{{\color{black}#1}}

\title{\rahul{Streamwise energy-transfer mechanisms in zero- and adverse-pressure-gradient turbulent boundary layers}}

\author{Rahul Deshpande\aff{1}
  \corresp{\email{raadeshpande@gmail.com}},
 and Ricardo Vinuesa\aff{2}}

\affiliation{\aff{1}Dept. Mechanical Engr., University of Melbourne, Parkville, VIC 3010, Australia
\aff{2}FLOW, Engr. Mechanics, KTH Royal Institute of Technology, Stockholm, 10044, Sweden}

\begin{document}
\maketitle

\begin{abstract}

\rahul{The present study investigates streamwise ($\overline{u^2}$) energy-transfer mechanisms in the inner and outer regions of turbulent boundary layers (TBLs).}
Particular focus is placed on the $\overline{u^2}$-production, its inter-component and wall-normal transport as well as dissipation, all of which become statistically significant in the outer region with increasing friction Reynolds number ($Re_{\tau}$).
\rahuld{These properties are analyzed using published data sets of zero, weak and moderately strong adverse-pressure-gradient (APG) TBLs across a decade of $Re_{\tau}$, revealing similarity in energy-transfer pathways for all these TBLs.}
It is found that both the inner and outer peaks of $\overline{u^2}$ are always associated with local maxima in the $\overline{u^2}$-production and its inter-component transport, and the regions below/above each of these peaks are always dominated by wall-ward/away-from-wall transport of $\overline{u^2}$, thereby classifying the $\overline{u^2}$-profiles into four distinct regimes.
This classification reveals existence of phenomenologically similar energy-transfer mechanisms in the `inner' and `outer' regions of moderately strong APG TBLs, which meet at an intermediate location coinciding with the minimum in $\overline{u^2}$ profiles.
\rahuld{Conditional averaging suggests existence of similar phenomena even in low-$Re_{\tau}$ canonical and/or weak APG TBLs, albeit with the outer-region mechanisms weaker than those in the inner region.} 
This explains the absence of their $\overline{u^2}$ outer peak and the dominance of $\overline{u^2}$ wall-normal transport away from the wall, which potentially originates from the inner region.
Given that the wall-ward/away-from-wall transport of $\overline{u^2}$ is governed by the $\rm Q_4$(sweeps)/$\rm Q_2$(ejections) quadrants of the Reynolds shear stress, it is argued that the emergence of the $\overline{u^2}$ outer peak corresponds with the statistical dominance of $\rm Q_4$ events in the outer region.
\rahul{Besides unravelling the dynamical significance of $\rm Q_2$ and $\rm Q_4$ events in the outer region of turbulent boundary layers, the present analysis also proposes new phenomenological arguments for testing on canonical wall-turbulence data at very high $Re_{\tau}$.}

\end{abstract}


\section{Introduction and motivation}
\label{intro}

The streamwise Reynolds stress ($\overline{u^2}$, where overbar denotes time or ensemble averaging) is one of the most widely analyzed statistical quantity in the wall-bounded turbulence literature.
This is not only because it is relatively convenient to measure, but also since it is the dominant contributor to the total turbulence fluctuations in any statistically two-dimensional wall flow.
The magnitude of $\overline{u^2}$, hence, is representative of the fundamental flow physics in different `regions' of a turbulent boundary layer (TBL).
For instance, $\overline{u^2}$ in the viscous (sublayer), inner and outer regions of the TBL is representative of the wall-shear stress fluctuations, dominance of the near-wall viscous-scaled streaks, and of the inertial large- and very-large-scale motions, respectively \citep{schoppa2002,baars2020part2}.
In very low-$Re_{\tau}$ zero pressure gradient (ZPG) TBLs, the near-wall streaks are the most prominent flow features \citep{kline1967}, leading to the appearance of a local maximum in $\overline{u^2}$ profiles at the location where they are centred (referred henceforth as the `inner peak').
Here, $Re_{\tau}$ = ${{\overline{U}}_{\tau}}{{\delta}_{99}}/{\nu}$ is the friction Reynolds number defined based on the mean friction velocity (${{\overline{U}}_{\tau}}$), 99\% TBL thickness (${\delta}_{99}$) and kinematic viscosity ($\nu$).

An increase in $Re_{\tau}$ leads to the growth in energy as well as broadening of hierarchy of the large- and very-large-scale motions, which originate in the outer region but have footprints extending down to the wall \citep{marusic2010high,baars2020part2}.
This results in the $Re_{\tau}$-growth of $\overline{u^2}$ magnitude across the TBL, but increasing most strongly in the outer region \citep{samie2018,baars2020part2}.
A longstanding question \citep{monkewitz2015,monkewitz2022} has been whether this $Re_{\tau}$ growth of $\overline{u^2}$ leads to the emergence of another local maximum farther away from the wall (referred henceforth as the `outer peak'), signifying dominance of the inertial large scales.
Investigating this, however, requires $Re_{\tau}$ increasing well beyond $10^4$ so that the integrated large-scale contributions are distinguishable from those of the small scales (\emph{i.e.} sufficient scale separation; \citealp{alfredsson2011}).
\rahul{The fact that very few experimental facilities can conduct accurate, well-resolved measurements in such high-$Re_{\tau}$ canonical flows \citep{fiorini2017,samie2018} has resulted in the discussion around the $\overline{u^2}$ outer peak still remaining an open question.}

On the other hand, imposition of an adverse pressure gradient (APG) on a TBL is well known to lead to a discernible outer peak in $\overline{u^2}$, even at low to moderate $Re_{\tau}$ \citep{pozuelo2022}.
Previous studies at low $Re_{\tau}$ have found this peak to be a consequence of the enhanced mean shear in an APG TBL, which significantly energizes the large-scale motions in the outer region via production of turbulent kinetic energy \citep{gungor2016,kitsios2017}.
\rahuld{While this changes the spatial organization and geometry of the coherent structures in low-to-moderately strong APG TBLs (compared to ZPG TBLs; \citealp{lee2009}), recent investigations have suggested that the dominant coherent structures \citep{lee2017} and their inherent energy-transfer mechanisms \citep{gungor2022} remain the same as in ZPG TBLs.
\citet{gungor2022} found that imposition of a weak-to-moderately strong pressure gradient predominantly changes only the intensity of the energy-transfer mechanisms in the inner and outer regions, relative to ZPG TBLs. 
Notably, even differences in the wall-attached/detached nature of the dynamically significant motions, between ZPG and APG TBLs, were found not to affect their energy-transfer mechanisms (see also discussion by \citealp{dong2017}).}
Inspired by these observations, the present study aims to understand the energy-transfer mechanisms associated with the $\overline{u^2}$ outer peak in weak-to-moderately strong APG TBLs, and assess whether similar mechanisms could be extended \rahul{to the outer region of high-$Re_{\tau}$ ZPG TBLs}.
This approach can potentially offer new flow-physics-based arguments \rahul{for future studies} supporting/against the emergence of the $\overline{u^2}$ outer peak in ZPG TBLs \rahul{(after availability of accurate, well-resolved measurements).}
Looking from a broader perspective, present approach could potentially provide an alternative pathway to understand the energy dynamics of very high $Re_{\tau}$ TBLs \citep{vinuesa2017wing}, which are omnipresent in engineering applications but hard to measure. 

\subsection{Streamwise Reynolds-stress transport}

While the standalone analysis of $\overline{u^2}$ has been extensively reported in the literature, its correlation with the individual terms in the transport equation of the streamwise turbulent kinetic energy (TKE) has been relatively limited owing to measurement challenges.
For a TBL with an imposed streamwise PG, the transport equation of streamwise TKE can be expressed as \citep{pozuelo2022,gungor2022}:
\begin{equation}\label{eq1}
{\frac{\partial}{{\partial}t}}{\overline{u^2}} = {{\mathcal{P}}^{u}} + {{\mathcal{E}}^{u}} + {{\mathcal{D}}^{u}} + {{\mathcal{T}}^{u}} \rahul{- {{\Pi}^{u}} + {{\Pi}^{u}_{t}}} + {{\mathcal{C}}^{u}},  
\end{equation}
where the production term, ${{\mathcal{P}}^{u}}$ = $-2{\overline{uw}}({{\partial}{\overline{U}}}/{{\partial}z})$ $-2{\overline{u^2}}({{\partial}{\overline{U}}}/{{\partial}x})$, the viscous dissipation ${{\mathcal{E}}^{u}}$ = 
$-2{\nu}{\sum_{j=1}^{3} {\overline{( {{{\partial}u}/{{\partial}{x_j}}} )^2}}}$, the viscous diffusion ${{\mathcal{D}}^{u}}$ = 
${\nu}{\sum_{j=1}^{3} ( {{{{\partial}^2}{\overline{u^2}}}/{{\partial}{x^2_j}}} )}$, \rahul{the pressure strain ${{\Pi}^{u}}$ = {$-({2}/{\rho})$}$\overline{p({{\partial}u}/{{\partial}x})}$, the pressure transport ${{\Pi}^{u}_{t}}$ = {$-({2}/{\rho})$}$({{\partial}\overline{pu}}/{{\partial}x})$}, the turbulent transport ${{\mathcal{T}}^{u}}$ = $-$(${{\partial}{\overline{{u^2}w}}}/{{\partial}z}$) and the convection ${{\mathcal{C}}^{u}}$ = {$-{\overline{U}}{{{\partial}{\overline{u^2}}}/{{\partial}x}}$} {$-{\overline{W}}{{{\partial}{\overline{u^2}}}/{{\partial}z}}$}.
Here, $u$, $v$ and $w$ represent turbulence fluctuations in streamwise ($x$), spanwise ($y$) and wall-normal ($z$) directions, respectively. 
Capital letters ($U$,$V$,$W$) represent instantaneous flow properties (\emph{i.e.} without subtracting the mean), while overbars (\emph{eg.} $\overline{U}$, $\overline{u^2}$) indicate time or ensemble averaging.
\citet{chen2021} have previously considered the energy balance in (\ref{eq1}) in the very near-wall region of a canonical wall-bounded flow, where ${{\mathcal{D}}^{u}}$ $\approx$ ${{\mathcal{E}}^{u}}$, to motivate the validity of the wall scaling of the $\overline{u^2}$ inner peak magnitude (found at $z_{\textrm{IP}}$) at $Re_{\tau}$ $\rightarrow$ $\infty$.
In the same region, \citet{tang2023} proposed a dissipation scaling for $\overline{u^2}$ profiles (\emph{i.e.}, based on ${{\mathcal{E}}^{j}}$ and $\nu$), which was found to extend across a larger wall-normal range with increasing $Re_{\tau}$.
\citet{monkewitz2015} estimated the scaling for the $\overline{u^2}$ outer peak (at $z_{\textrm{OP}}$) in a canonical flow, based on a simplified balance between the net turbulence production and dissipation, yielding $z^+_{\textrm{OP}}$ $\approx$ ${({0.06}\:{{\mathcal{E}}^{j}})}^{-1}$, where $j$ = 1--3.
However, it is now well known that the imbalance between ${{\mathcal{P}}^{j}}$ and ${{\mathcal{E}}^{j}}$ in the outer region increases with $Re_{\tau}$, which is accounted for by the wall-normal transport, ${{\mathcal{T}}^{j}}$ \citep{cho2018,mklee2019}. 
Considering that our main goal is to understand the high-$Re_{\tau}$ dynamics in the outer region, the present study focuses on correlating the wall-normal profiles of relevant $Re_{\tau}$-dependent terms in (\ref{eq1}) with $\overline{u^2}$. 

In a canonical wall flow, the increase in ${\mathcal{P}}^u$ (\emph{i.e.}, the TKE source term) is associated with the broadening and energization of the inertial eddy hierarchy with increasing $Re_{\tau}$ \citep{marusic2010high,baidya2017}, which is accompanied by an increase in energy dissipation
(${\mathcal{E}}^u$, \emph{i.e.} the sink term) predominantly via the small dissipative scales \citep{mklee2019}.
In terms of variations across the wall-normal direction, the energy pathway from the source to sink comprises the pressure strain term ${\Pi}^u$, representing transfer from $\overline{u^2}$ to $\overline{v^2}$ and $\overline{w^2}$, and the wall-normal transport term ${{\mathcal{T}}^{u}}$ = ${{\partial}}({\overline{{u^2}w}})/{{\partial}{z}}$, representing diffusion of $\overline{u^2}$ from the outer region. 
Both of these are known to increase with $Re_{\tau}$ alongside ${\mathcal{P}}^u$ and ${\mathcal{E}}^u$ \citep{mklee2019}. 
Although $\overline{u^2}$ is dependent on the net balance of all the terms in equation (\ref{eq1}), the above discussion points towards an increasing influence of ${\mathcal{P}}^{u}$, ${\mathcal{T}}^{u}$, ${\Pi}^u$ and ${{\mathcal{E}}^{u}}$ on $\overline{u^2}$ with increasing $Re_{\tau}$, especially in the outer region.
Looking further into the definition of ${\mathcal{T}}^{u}$, the triple product represents net flux of $\overline{u^2}$-energy transported in the wall-normal direction, with $\overline{{u^2}w}$ $<$ 0 and $\overline{{u^2}w}$ $>$ 0 respectively indicating flux towards or away from the wall \citep{mklee2019}.
In the case of relatively low $Re_{\tau}$ canonical flows, \citet{mklee2019} noted that the mean energy transport in the outer region is directed away from the wall (\emph{i.e.} $\overline{{u^2}w}$ $>$ 0).
Based on their $Re_{\tau}$ trends, however, \citet{mklee2019} hypothesized that significant increase in ${\mathcal{P}}^{u}$ at high $Re_{\tau}$ would be accompanied with mean energy flux from the intermediate region towards the wall (\emph{i.e.} $\overline{{u^2}w}$ $<$ 0).
This suggests increasing dominance of the streamwise momentum-carrying motions moving towards the wall (\emph{i.e.} $\rm Q_4$ events: $u$ $>$ 0, $w$ $<$ 0; \citealp{deshpande2021uw}), in very-high-$Re_{\tau}$ flows where a $\overline{u^2}$ outer peak may emerge.

Interestingly, analogous trends have been noted in the literature for the case of moderately strong APG TBLs, where a $\overline{u^2}$ outer peak is clearly observed \citep{skaare1994turbulent,lee2017}.
Both these studies found the outer-peak region in ${{\mathcal{P}}^{u}}$ and $\overline{u^2}$ to coincide with dominance of $\overline{u^2}$ energy flux towards the wall (\emph{i.e.} $\overline{{u^2}w}$ $<$ 0), with the latter driven by significantly energized $\rm Q_4$ events.
\citet{lee2017} used conditional analysis to confirm intensification of these $\rm Q_4$ events is governed by the large-scale roll modes in the APG TBL outer region.
Besides these speculated similarities in ${{\mathcal{T}}^{u}}$, \citet{gungor2022} recently conducted two-dimensional spectral analysis of ${\mathcal{P}}^u$ and ${\Pi}^u$ to demonstrate similarities in these transfer mechanisms for canonical flows and moderately strong APG TBLs.
Hence, \rahuld{although the dominant mechanisms behind enhancement of the TKE source term (${{\mathcal{P}}^{u}}$) are different for APG and ZPG TBLs (at least at low $Re_{\tau}$), the energy-transfer pathways between ${{\mathcal{P}}^{u}}$, ${\Pi}^u$ and ${{\mathcal{T}}^{u}}$, on energization of APG TBL outer region, seem analogous to that hypothesized for very-high-$Re_{\tau}$ ZPG TBLs.
This reaffirms the motivation for the present investigation.}

\subsection{\rahul{Present contributions}}

The present study analyzes the energy-transfer mechanisms behind the $\overline{u^2}$ outer peak using published ZPG and APG TBL data sets at various matched $Re_{\tau}$ (described in $\S$\ref{setup}).
\rahuld{The first analysis in $\S$\ref{results1} compares all the dominant terms of $\overline{u^2}$-transport with respect to increasing $Re_{\tau}$, \emph{i.e.} ${\mathcal{P}}^u$, ${\mathcal{E}}^u$, ${\mathcal{T}}^u$ and ${\Pi}^u$.
It confirms the similarity in energy-transfer pathways for ZPG and APG TBLs, across a larger $Re_{\tau}$ range than that considered by \citet{gungor2022}.
This result motivates $\S$\ref{results1a}, which reveals the increasing dominance of $\overline{uw}$ in TKE production with increasing $Re_{\tau}$, for both ZPG and weak-to-moderately strong APG TBLs.}
Given $uw$ influences both ${\mathcal{P}}^{u}$ and ${\mathcal{T}}^{u}$, a comprehensive analysis of the $uw$-contributions from the individual $\rm Q_2$ (ejections; $u$ $<$ 0, $w$ $>$ 0) and $\rm Q_4$ (sweeps; $u$ $>$ 0, $w$ $<$ 0) events, relative to the streamwise TKE, is discussed in $\S$\ref{results2}. 
These two quadrants are the most dominant out of the four possible quadrants of the Reynolds shear stress in a wall-bounded flow ($Q_i$, with $i$ = 1--4; \citealp{wallace2016}).
While both $\rm Q_2$ and $\rm Q_4$ events contribute positively to the TKE production, the opposite signs in $w$-fluctuations imply that they are individually responsible \citep{nagano1990} for $\overline{u^2}$ energy flux towards ($\rm Q_4$, $\overline{{u^2}w}$ $<$ 0) and away from the wall ($\rm Q_2$, $\overline{{u^2}w}$ $>$ 0).
Hence, investigating variation in their contributions can potentially explain changes in the direction of $\overline{u^2}$ flux from the outer region, which is noted on increasing APG strength \citep{skaare1994turbulent} or increasing $Re_{\tau}$ \citep{mklee2019,deshpande2021uw}.
To the authors' knowledge, the present study is the first attempt of using the wall-normal profiles of ${uw}$-quadrant contributions to understand the energy dynamics around the $\overline{u^2}$ peaks \citep{wallace2016}.

\begin{table}
   \captionsetup{width=1.0\linewidth}
   \centering
  \begin{center} 
    \begin{tabular}{llllllll}
      Data set & TBL & $Re_{\tau}$ & $\beta$ & \# of & Color & Resolution & Reference \\
       &  &  &  & cases & shading & (${{\Delta}{y^+_{m}}}{\times}{{\Delta}{z^+_{m}}}$) \\    
      \noalign{\smallskip}\hline\noalign{\smallskip}
    \vspace{1.5mm}
      LES & ZPG & 290-2000 & 0 & 7 & Black & 10.9 $\times$ 19.6 & \citet{eitel2014}\\     
      \vspace{1.5mm}
      LES & APG & 300-750 & 0.85-1.0 & 3 & {\color{violet}Violet} & 9.7 $\times$ 8.4 & \citet{bobke2017}\\
      \vspace{1.5mm}
      LES & APG & 450-2000 & 0.85-1.6 & 6 & {\color{red}Red} & 12.5 $\times$ 30.1 & \citet{pozuelo2022}\\
      \multirow{2}{*}{Exp.} &
      \multirow{2}{*}{APG} & 
      1900-4000 & 1.3 & 2 & {\color{golden}Golden} & 20 $\times$ 100 & \citet{sanmiguel2017}\\
      && 1000-4000 & 2.4 & 3 & {\color{magenta}Magenta} & 20 $\times$ 100 & \citet{sanmiguel2017}\\      
    \end{tabular}
  \end{center}
    \caption{Table summarizing the parametric space associated with various published experimental (Exp.) and LES data sets analyzed in this study. 
    Definitions/terminologies have been provided in either $\S$\ref{intro} or \ref{setup}. 
    Throughout this manuscript, light to dark color shading indicates increasing $Re_{\tau}$. 
    The pressure-gradient history associated with each case can be visualized in figure \ref{fig1}. 
    In case of LES, ${\Delta}{y_m}$ and ${\Delta}{z_m}$ refer to the numerical grid resolution in the free-stream, while it refers to the interrogation window size in case of Exp. data sets.}   
  \label{tab1}
\end{table}

\section{Experimental and numerical data sets}
\label{setup}

We consider \rahul{four} previously published experimental and numerical data sets for the present analysis, covering both ZPG and APG TBLs across 290 $\lesssim$ $Re_{\tau}$ $\lesssim$ \rahul{4000}.
Three of these data sets are from the high-resolution large-eddy-simulations (LES) of a ZPG TBL \citep{eitel2014} and APG TBLs \citep{bobke2017,pozuelo2022}. 
\rahul{The fourth is an} experimental data set acquired via planar particle image velocimetry (PIV; \citealp{sanmiguel2017}) in a APG TBL.
The high-resolution-LES data considered here has a resolution two times coarser than a typical DNS in the streamwise and spanwise directions. 
A small body force is added to account for the unresolved TKE dissipation (around 10\% of the total), results from which have been shown to agree extremely well with DNS in both canonical \citep{eitel2014} and more complex \citep{negi2018} turbulent flows. 
The only effect of the lower resolution is a small attenuation of the near-wall $\overline{u^2}$ peak, which however does not influence the present discussion.
Our analysis only depends on identification of the $\overline{u^2}$ peak and its $z$-location, which is not affected by use of high-resolution LES (as compared to DNS).

\begin{figure}
   \captionsetup{width=1.0\linewidth}
\centering
\includegraphics[width=1.0\textwidth]{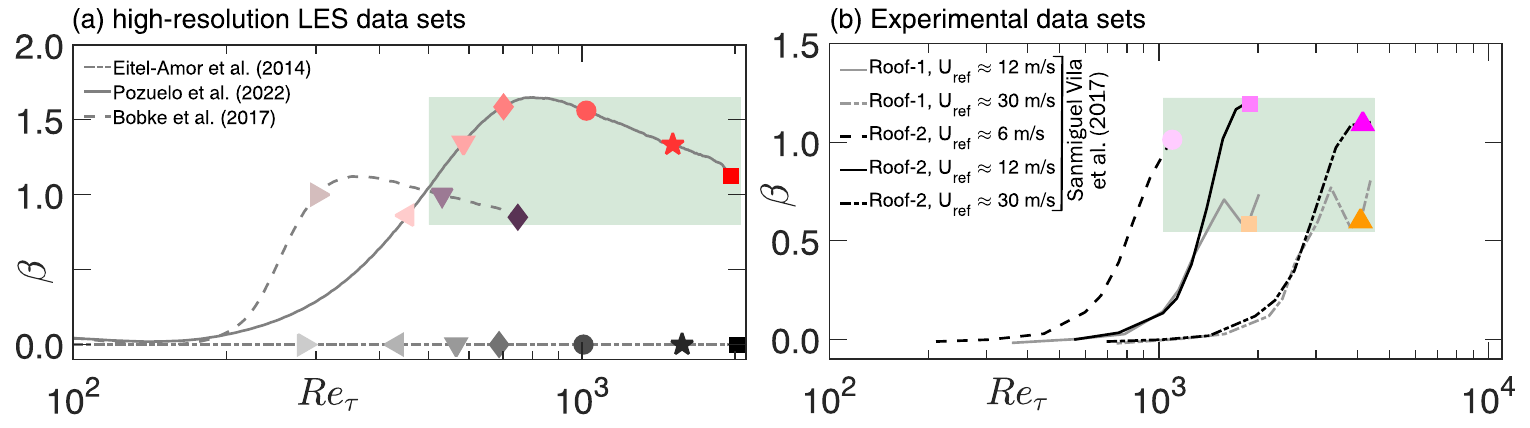}
\caption{$\beta$ distribution as a function of $Re_{\tau}$ corresponding to the (a) high-resolution LES and (b) experimental data sets analyzed in the present study and documented in table \ref{tab1}.
Lines represent pressure-gradient histories across the (a) computational domains and (b) wind tunnels associated with these data sets, while symbols represent selected cases for which the turbulence statistics are analyzed.
Light to dark shading in symbol colors represent increasing $Re_{\tau}$, with cases at matched $Re_{\tau}$ (across different data sets) represented by the same symbols.
Green background shading is used to highlight cases where an outer peak is noted in the $\overline{u^2}$ profiles.}
\label{fig1}
\end{figure}

For all the data sets, the boundary-layer thickness (${\delta}_{99}$) is estimated via the diagnostic-plot method (see \citealp{vinuesa2016}). 
The same definition of ${\delta}_{99}$ has also been used to calculate the Clauser-pressure gradient parameter, ${\beta}$ $=$ $(\delta^* / {\rho {\overline{U}}_\tau^2}) \left( {\rm d} \overline{P}/ {\rm d} x \right)$ \citep{bobke2017}, where $\delta^* = \int_0^{\delta_{99}}(1 - {\overline{U}}(z)/{\overline{U_e}})\:{\rm d}z$ is the displacement thickness,  
${\overline{U_e}}$ $=$ ${\overline{U}}$($z$ = $\delta_{99}$), \textit{i.e.} the edge velocity, $\rho$ is the fluid density and ${\rm d} {\overline{P}}/{\rm d} x$ is the mean streamwise pressure-gradient at the $x$-location where $\beta$ is estimated.
All the data sets have been tabulated in table \ref{tab1}, while their $\beta$ distribution as a function of $Re_{\tau}$ (across the experimental/LES domains) is shown in figure \ref{fig1} to provide a general overview.

As can be noted from figure \ref{fig1}(a), a unique aspect of the LES data sets of APG TBL is their near-constant $\beta$ distribution across a substantial portion of the computational domain.
In the case of \citet{bobke2017}, $\beta$ $\approx$ 1 for 300 $\lesssim$ $Re_{\tau}$ $\lesssim$ 750, while in the case of \citet{pozuelo2022}, $\beta$ $\approx$ 1.4 for 700 $\lesssim$ $Re_{\tau}$ $\lesssim$ 2000. 
Subsequently, when combined together, they facilitate investigation of the variation of the flow physics in a near-equilibrium TBL across a broad $Re_{\tau}$ range.
Despite the near-constant $\beta$, however, the statistics would also be dependent on the unique upstream pressure gradient history, across $Re_{\tau}$ $\lesssim$ 300 \citep{bobke2017} and $Re_{\tau}$ $\lesssim$ 700 \citep{pozuelo2022}.
Thus, in order to demonstrate that the present conclusions are independent of the upstream history effects, this study analyzes turbulence statistics in both nominally constant $\beta$ as well as varying $\beta$ sections of the computational domain.

Wall-normal variation of the turbulence statistics is considered at selected $x$-locations of the computational domain, each corresponding to a unique $Re_{\tau}$, which have been indicated with symbols in figure \ref{fig1}(a).
All these selected points are sufficiently far away from the start of the computational domain (refer to \citealp{bobke2017} and \citealp{pozuelo2022}).
A reference data set for ZPG TBL \citep{eitel2014}, spanning a $Re_{\tau}$-range similar to that covered by the two APG TBL LES data sets, is also considered. 
Turbulence statistics are analyzed at seven streamwise locations within the ZPG domain, such that their corresponding $Re_{\tau}$ match reasonably well with the selected three and six cases from the APG domains of \citet{bobke2017} and \citet{pozuelo2022}, respectively. 
For convenience in referencing, these cases at matched $Re_{\tau}$ will be represented by the same symbols in the forthcoming analysis (figure \ref{fig1}a).
The turbulence statistics presented here are computed by averaging across the time series saved at these selected $x$-positions, as well as along the homogeneous spanwise direction.
The computational grid resolution (table \ref{tab1}) has been demonstrated in the original references to be fine enough to ensure that the peaks/maxima in the $\overline{u^2}$ profiles are physical, and not an artefact of spatial-resolution issues \citep{samie2018}.
It is worth noting here that the (${\Delta}{y^+_{m}}$,${\Delta}{z^+_{m}}$)-values in table \ref{tab1} refer to the maximum spanwise and wall-normal grid spacing. 
The grid is very fine close to the wall, with 7 points below $z^+$ $<$ 10.

The experimental data set for APG TBL was acquired via two-component two-dimensional (2-D) PIV in the streamwise-wall normal ($x$-$z$) plane at the Minimum Turbulence Level (MTL) facility at KTH, Stockholm \citep{sanmiguel2017}.
The pressure gradient was imposed by hanging wall inserts in a converging-diverging geometry from the tunnel roof.
The converging section initially led to the imposition of a favorable pressure gradient, followed by a region of nominal ZPG (in the straight section) and finally a region of APG (in the diverging section), making its upstream history different than that noted for the high-resolution LES cases.
The turbulent boundary layer was measured over a smooth aluminium plate, which was suspended $\sim$15\:cm from the tunnel floor and had its leading edge (\emph{i.e.} $x$ $\approx$ 0) upstream of the ZPG region.
Overall, five different $\beta$ distributions with unique upstream histories (figure \ref{fig1}b) were obtained by using two different roof geometries (indicated by different colors) and three different upstream free-stream speeds of $U_{\rm ref}$\:$\sim$\:6, 12 and 30\: m/s (indicated by color shading intensity).
Here, $U_{\rm ref}$ is the free-stream speed at $x$ $\approx$ 0.
2-D PIV was conducted sufficiently downstream ($\sim$4.5\:m) from the leading edge of the plate, such that a moderately strong APG (1.3 $\lesssim$ $\beta$ $\lesssim$ 2.4) was imposed on the TBL.
Indeed, the spatial resolution of these PIV experiments is much lower than those of the other simulations and experiments considered (table \ref{tab1}). 
However, the $\overline{u^2}$ and $\overline{uw}$ profiles obtained from these measurements have been demonstrated by \citet{sanmiguel2017} to be sufficiently well-resolved in the outer-region ($z^+$ $\gtrsim$ 100) of the APG TBL, and we will limit our investigation of this data set only to this region.
In their original paper, the authors also employed two post-processing approaches on the same PIV data set to further improve the accuracy of the near-wall turbulence statistics, which however are not considered here.

\section{\rahuld{Similarity in energy-transfer pathways for ZPG and APG TBLs}}
\label{results1}

\begin{figure}
   \captionsetup{width=1.0\linewidth}
\centering
\includegraphics[width=0.97\textwidth]{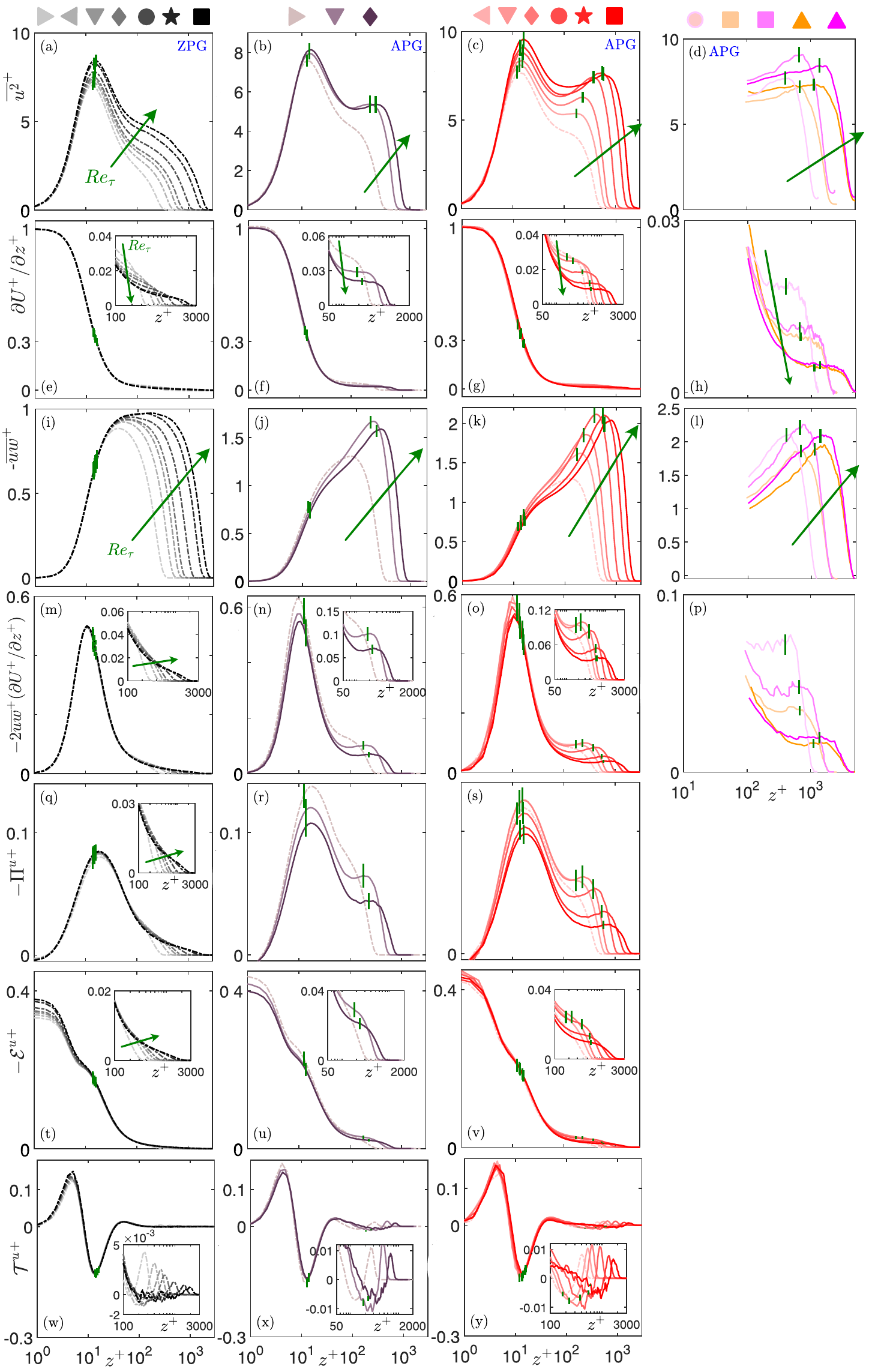}
\caption{Wall-normal profiles for (a-d) ${\overline{u^2}}^{+}$, (e-h) ${{\partial}{U^+}/{\partial}{z^+}}$, (i-l) $-{{\overline{uw}}^{+}}$, (m-p)  ${{\mathcal{P}}^{u}}^+$ $\approx$ $-2{{\overline{uw}}^{+}}{{\partial}{U^+}/{\partial}{z^+}}$, (q-s) $-{{\Pi}^{u}}^{+}$, (t-v) $-{{\mathcal{E}}^{u}}^{+}$ and (w-y) ${{\mathcal{T}}^{u}}^{+}$ for ZPG (left column) and APG TBL cases (other columns).
Symbols/colors correspond to the various cases defined in figure \ref{fig1}, with light to dark shading indicating increasing $Re_{\tau}$.
Note the inconsistent axes range.}
\label{fig2}
\end{figure}

We begin our analysis by connecting the $\overline{u^2}$ profiles from all the data sets (figure \ref{fig2}a-d), with their corresponding terms in the streamwise Reynolds-stress transport equation: production (${{\mathcal{P}}^{u}}^+$; figures \ref{fig2}m-p), pressure-strain (${{{\Pi}}^{u}}^+$; figures \ref{fig2}q-s), viscous dissipation (${{\mathcal{E}}^{u}}^+$; figures \ref{fig2}t-v) and wall-normal turbulent transport (${{\mathcal{T}}^{u}}^+$; figures \ref{fig2}w-y).
\rahuld{The intention is to compare the energy-transfer pathways, starting from the TKE source ${\mathcal{P}}^u$, for both ZPG and APG TBLs.}
Here, we only consider the dominant term for streamwise TKE production: ${\mathcal{P}}^{u}$ $\approx$ $-$2${\overline{uw}}({{\partial}{\overline{U}}}/{{\partial}{z}})$, and neglect the other term included previously in (\ref{eq1}), $-$2${\overline{u^2}}({{\partial}{\overline{U}}}/{{\partial}{x}})$, considering its negligible contribution \citep{pozuelo2022}.
While ${{\mathcal{P}}^{u}}^+$ could be obtained from the PIV dataset of APG TBL \citep{sanmiguel2017}, other properties such as ${{\mathcal{E}}^{u}}^+$, ${{\mathcal{T}}^{u}}^+$ and ${{\Pi}^{u}}^+$ were either not possible to estimate, or were erroneous owing to their sensitivity to the wall-normal gradient.
In figure \ref{fig2}, the locations of the inner ($z^{+}_{\textrm{IP}}$) and outer peaks ($z^+_{\textrm{OP}}$) of $\overline{u^2}$ were identified by estimating the gradients of the profiles, \emph{i.e.} ${\textrm{d}}{\overline{u^2}}$/${\textrm{d}}z$ = 0 and ${{\textrm{d}}^2}{\overline{u^2}}$/${\textrm{d}}{z^2}$ $<$ 0, and the locations of both the peaks are indicated by vertical green lines across all the plots in the figure.

The inner peak of $\overline{u^2}$ for both ZPG and APG TBLs is close to the local maximum of ${{\mathcal{P}}^{u}}^+$ observed in the near-wall region. 
This close correspondence is consistent with previous observations \citep{kline1967,mklee2019} and is associated with the intense TKE production by the streaks related to the near-wall cycle \citep{hamilton1995}.
\rahuld{The significant production of TKE at $z^+_{\rm IP}$ is connected with a local maximum of --${{\Pi}^{u}}^+$ and a local minima in ${{\mathcal{T}}^{u}}^+$, for both ZPG and APG TBLs, suggesting similar energy-transfer pathways in their respective inner regions (noted previously for ${{\mathcal{P}}^{u}}^+$ and --${{\Pi}^{u}}^+$ by \citealp{gungor2022}).}
Here, the loss in ${\overline{u^2}}$ via ${{\Pi}^{u}}^+$ acts as the gain term for the lateral velocity components ${\overline{v^2}}$ and ${\overline{w^2}}$ via ${{\Pi}^{v}}^+$ and ${{\Pi}^{w}}^+$, respectively.
Similarly, a negative value at the minima of ${{\mathcal{T}}^{u}}^+$ represents wall-normal transport of ${\overline{u^2}}$ from $z^+_{\rm IP}$ to other regions of the TBL \citep{mklee2019}.

\begin{figure}
   \captionsetup{width=1.0\linewidth}
\centering
\includegraphics[width=0.85\textwidth]{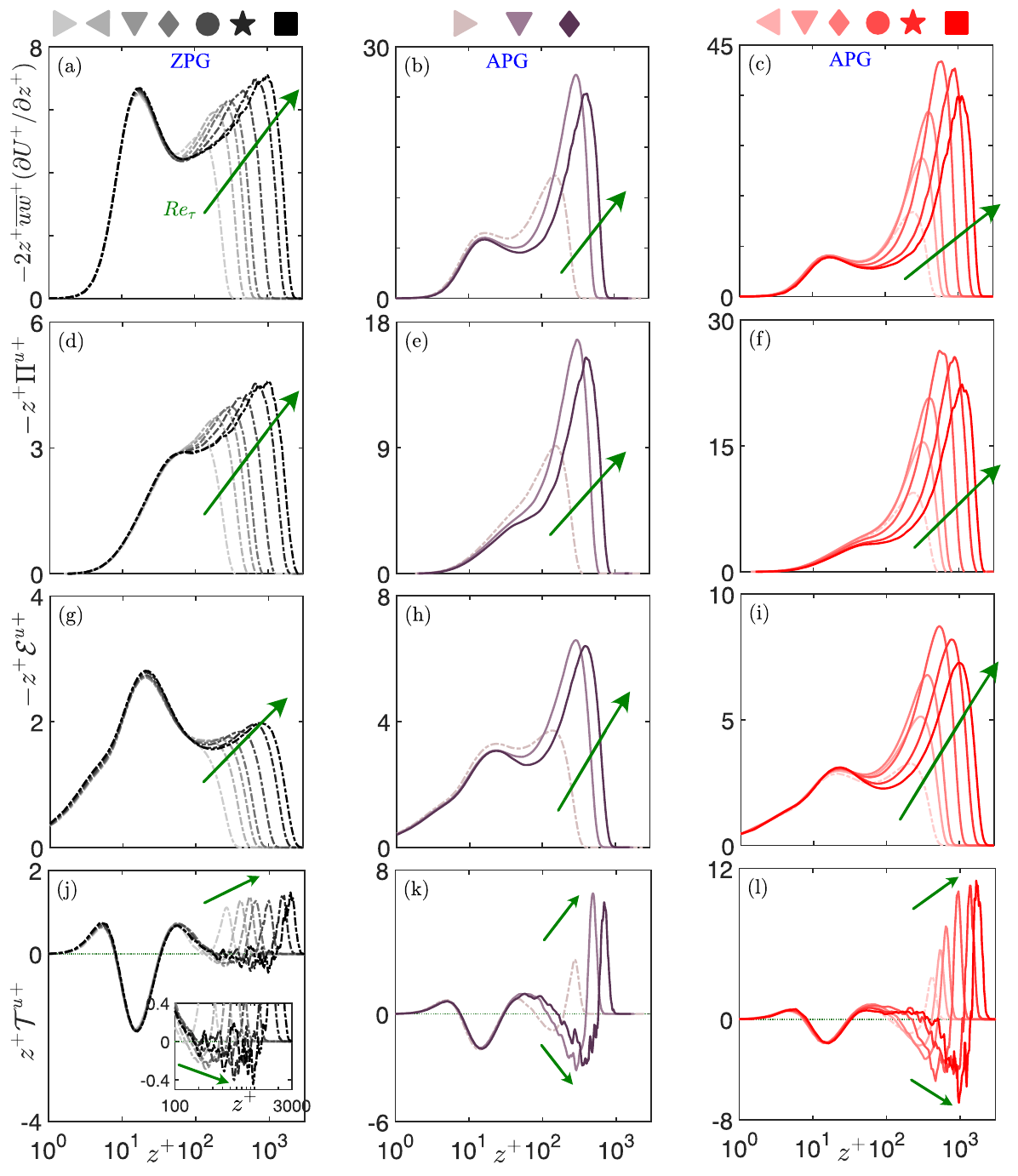}
\caption{Wall-normal profiles for $\overline{u^2}$-transport terms premultiplied by $z^+$: (a-c) ${z^+}{{\mathcal{P}}^{u}}^+$ $\approx$ $-2{z^+}{{\overline{uw}}^{+}}{{\partial}{U^+}/{\partial}{z^+}}$, (d-f) $-{z^+}{{\Pi}^{u}}^{+}$, (g-i) $-{z^+}{{\mathcal{E}}^{u}}^{+}$ and (j-l) ${z^+}{{\mathcal{T}}^{u}}^{+}$ for ZPG (left column) and APG TBL cases.
Symbols/colors correspond to the various cases defined in figure \ref{fig1}, with light to dark shading indicating increasing $Re_{\tau}$.
Note the inconsistent range of ordinate.}
\label{fig3}
\end{figure}

While the ${{\mathcal{P}}^{u}}^+$ profiles for ZPG TBL seem to collapse reasonably well in inner-scaling for a major portion of the TBL, that is not the case in the far outer region (figure \ref{fig2}m).
\rahuld{In fact, the bulk TKE production (\emph{i.e.} ${z^+}{{\mathcal{P}}^{u+}}$; figure \ref{fig3}a) in the outer region increases with $Re_{\tau}$, thereby driving the $Re_{\tau}$ growth of $\overline{u^2}$ evident in figure \ref{fig2}(a).
This increase in ${{\mathcal{P}}^{u}}^+$ leads to an increase in magnitude of ${{\mathcal{T}}^{u}}^+$, as well as enhanced losses in ${\overline{u^2}}$ via ${{\mathcal{E}}^{u}}^+$ and ${{\Pi}^{u}}^+$ with $Re_{\tau}$ \citep{mklee2019}, which can also be noted from figures \ref{fig3}(d,g,j) depicting transport terms premultiplied by $z^+$.}
\rahul{While such a qualitative comparison of the premultiplied transport terms is reasonable, due precaution is required when interpreting them quantitatively.
Premultiplication with $z^+$ artificially dampens the energy in the inner region while exaggerating that in the outer region, an example of which is evident from the `outer peak' in ${z^+}{{\mathcal{P}}^u}^+$ (figure \ref{fig3}a) which however does not exist in figure \ref{fig2}(m).}

The outer region of APG TBLs exhibit significantly enhanced ${{\mathcal{P}}^{u}}^+$ even at low $Re_{\tau}$ (figure \ref{fig2}n-o), albeit due to a different mechanism than for high-$Re_{\tau}$ ZPG TBLs.
This enhanced ${{\mathcal{P}}^{u}}^+$ is again accompanied with significantly high magnitudes of ${{\mathcal{T}}^{u}}^+$ as well as greater losses in ${\overline{u^2}}$ via ${{\mathcal{E}}^{u}}^+$ and ${{\Pi}^{u}}^+$, when compared to ZPG TBLs at matched $Re_{\tau}$.
\rahuld{Qualitatively, the energization of the outer region in APG TBLs is consistent with that exhibited by ZPG TBLs on increasing $Re_{\tau}$,} \rahul{except the fact that statistics for moderately-strong APG TBLs exhibit a distinct maximum (owing to greater energization of the outer region)}.
\rahuld{Notably, this similarity in the energy-transfer mechanisms of the outer region has also been discussed previously by \citet{gungor2022}, but is confirmed for all major Reynolds stress transport terms in figures \ref{fig2} and \ref{fig3}.}
\citet{harun2013}, \citet{pozuelo2022} and \citet{gungor2022} have also confirmed that this energy enhancement is not an arterfact of use of inner scaling (\emph{i.e.}, with ${\overline{U}}_{\tau}$), but is physical and associated with energization of the outer region on imposition of APG.

\rahuld{Interestingly, analogous to the energy-transfer mechanisms noted near the inner peak, there is a local maximum in ${{\mathcal{P}}^{u}}^+$ as well as --${{\Pi}^{u}}^+$ at the ${\overline{u^2}}$ outer peak location, $z^+_{\rm OP}$ (see vertical green lines in figure \ref{fig2}b-d,n-p,r-s).
The outer peak is also close to the minima in ${{\mathcal{T}}^{u}}^+$ which is again negative (figure \ref{fig2}x,y), while no maxima is noted in --${{\mathcal{E}}^{u}}^+$  (figure \ref{fig2}u,v).
On the other hand, there is no local maximum in ${{\mathcal{P}}^{u}}^+$ and --${{\Pi}^{u}}^+$ for scenarios without an outer peak of $\overline{u^2}$ (see dash-dotted profiles), reaffirming the one-to-one connection between these statistics.}
Hence, similar to the scenario noted at $z^+_{\rm IP}$, the significant TKE production at $z^+_{\rm OP}$ is also accompanied by enhanced energy transfer from $\overline{u^2}$ to $\overline{v^2}$ and $\overline{w^2}$ (via the pressure-strain mechanism), and the wall-normal transport of $\overline{u^2}$ to other regions of the APG TBL.
\rahul{Future work would focus on investigating these statistics for very-high-$Re_{\tau}$ canonical flow data (where a $\overline{u^2}$ outer peak may emerge), to check for quantitative similarity with moderately-strong APG TBLs (having significantly enhanced outer region energy).
However, considering the paucity of such data in today's date, we employ conditional analysis on relatively low-$Re_{\tau}$ ZPG TBL data in $\S$\ref{results2b}, to unravel the qualitative similarity between ZPG and APG TBLs.}
Further, we present elaborate analyses on the similarity in the wall-normal energy transport, between the TBL inner and outer regions, in $\S$\ref{results2a} and $\S$\ref{results2b}.
However, first we focus on understanding the $Re_{\tau}$-variations in ${{\mathcal{P}}^{u}}^+$ for ZPG and APG TBLs in $\S$\ref{results1a}, given it is the source of the streamwise TKE that leads to both the $\overline{u^2}$ peaks \citep{harun2013,gungor2016,kitsios2017}.

\subsection{Assessing the enhancement of TKE production with Reynolds number}
\label{results1a}

\begin{figure}
   \captionsetup{width=1.0\linewidth}
\centering
\includegraphics[width=0.85\textwidth]{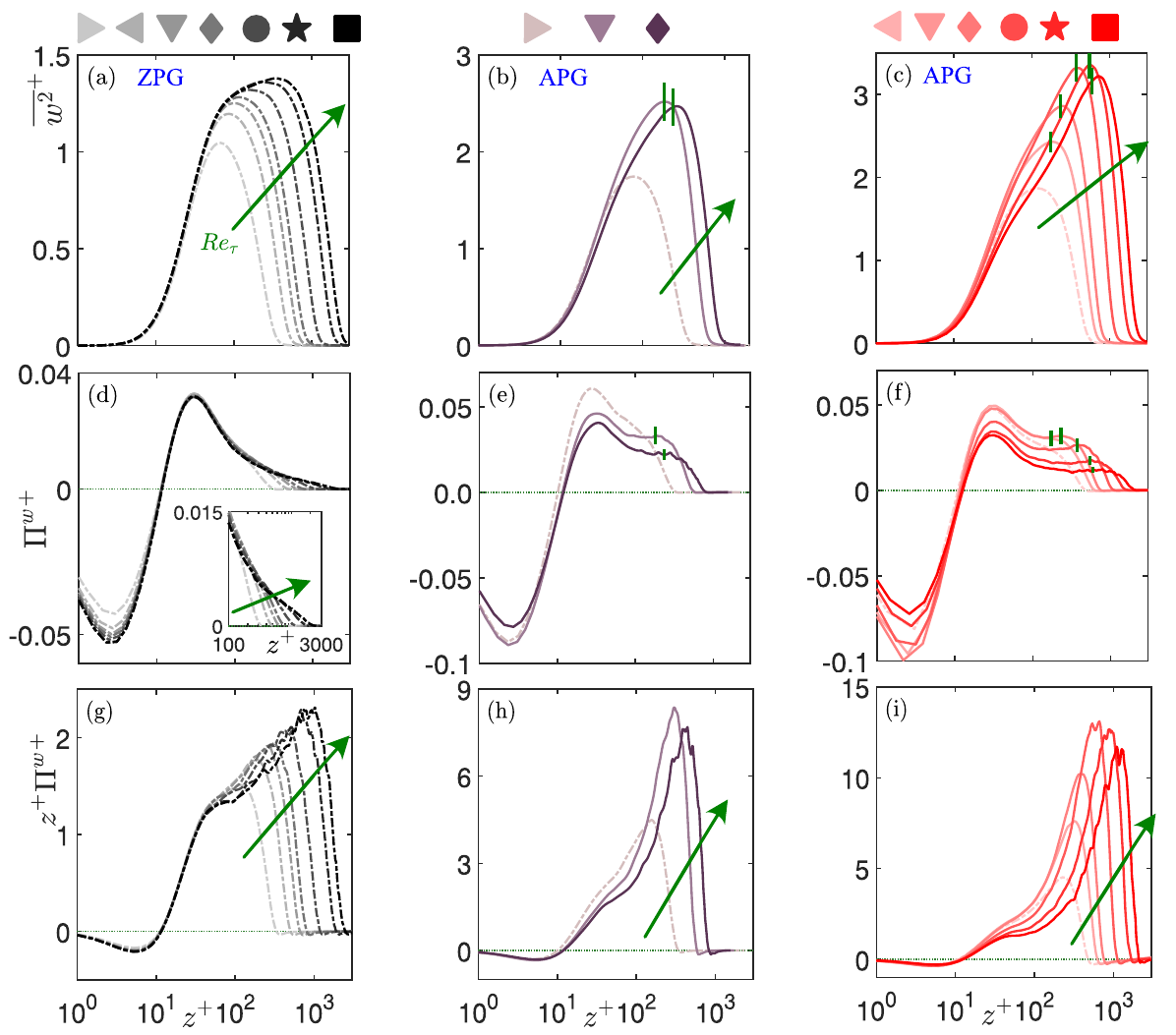}
\caption{Wall-normal profiles for (a-c) ${\overline{w^2}}^+$, (d-f) ${{\Pi}^{w}}^{+}$ and (g-i) ${z^+}{{\Pi}^{w}}^{+}$ for ZPG (left column) and APG TBL cases.
Symbols/colors correspond to the various cases defined in figure \ref{fig1}, with light to dark shading indicating increasing $Re_{\tau}$.
Note the inconsistent range of ordinate.
Vertical green lines indicate $\overline{u^2}$ outer peak locations.}
\label{fig4}
\end{figure}

Considering its mathematical definition, variations in ${{\mathcal{P}}^{u}}^+$ 
can be caused by either changes in the mean shear (figure \ref{fig2}e-h), magnitude of the Reynolds shear stress ($\overline{uw}$; figure \ref{fig2}i-l) or a combination of both. 
The mean shear is highest in the near-wall region of both ZPG and weak-to-moderate APG TBLs, and hence is predominantly responsible for their inner peak in ${{\mathcal{P}}^{u}}^+$ and $\overline{u^2}$ \citep{kline1967}.
However, the mean shear is significantly low in the outer region of ZPG TBLs and reduces gradually with increasing $Re_{\tau}$ (figure \ref{fig2}e).
The increase in bulk production with $Re_{\tau}$, hence, can be exclusively associated with the gradual increase and broadening of the quasi-constant $\overline{uw}$ plateau (figure \ref{fig2}i) predominantly around the overlap region of the ZPG TBL (\emph{i.e.}, a subset of the outer region). 
This has been extensively demonstrated via both experimental and numerical data in the literature \citep{marusic2010high,baidya2017}, and is associated with the energization and broadening of the inertia-dominated eddy hierarchy with increasing $Re_{\tau}$.
\rahuld{Here, the increase in $\overline{uw}$ can be connected with the increase in magnitude of both $u$- as well as $w$-fluctuations in the overlap region, and is depicted for the latter by ${\overline{w^2}}^+$ plotted for ZPG TBLs in figure \ref{fig4}(a).
This growth can be traced back to the source term for ${\overline{w^2}}$, \emph{i.e.} ${{\Pi}^{w}}^+$ (figure \ref{fig4}d), which also clearly increases in the outer region with $Re_{\tau}$ owing to the increased losses from $\overline{u^2}$, via ${{\Pi}^{u}}^+$ (also refer ${z^+}{{\Pi}^{u}}^+$, ${z^+}{{\Pi}^{w}}^+$ in figures \ref{fig3}d and \ref{fig4}g).}

In the case of low-$Re_{\tau}$ APG TBLs, interestingly, the enhanced ${{\mathcal{P}}^{u}}^+$ in the outer region (relative to ZPG TBLs) can be associated with relatively high mean shear (figure \ref{fig2}e-h) as well as $\overline{uw}$ (figure \ref{fig2}i-l).
\rahuld{However, an increase in $Re_{\tau}$ results in a drop in the mean shear in their outer region, and hence the sustained high magnitudes and peaks of ${{\mathcal{P}}^{u}}^+$ at high $Re_{\tau}$ can be associated with the increasing $\overline{uw}$ magnitudes noted for low-to-moderately strong APG TBLs.
This scenario is consistent with the $Re_{\tau}$-trends noted in case of ZPG TBLs,
which is also confirmed from the high magnitudes of $w$-fluctuations (figure \ref{fig4}b,c) sourced from the significantly enhanced ${{\Pi}^{w}}^+$ and $-{{\Pi}^{u}}^+$ in the outer region (figures \ref{fig2}-\ref{fig4}).}
Indeed, one can note a clear one-to-one connection between ${{\mathcal{P}}^{u}}^+$, --${{\Pi}^{u}}^+$, ${{\Pi}^{w}}^+$, ${\overline{w^2}}^+$ and --${\overline{uw}}^+$ profiles for APG TBLs in figures \ref{fig2} and \ref{fig4}, which is reflected by the close alignment of their respective outer peaks with $z^+_{\rm OP}$, as also noted previously in high-$Re_{\tau}$ APG TBLs \citep{romero2022properties}.
Consistent $Re_{\tau}$-trends are also exhibited by the bulk energy-transfer terms in the outer region (figures \ref{fig3} and \ref{fig4}), corresponding to the transport of both $\overline{u^2}$ and  $\overline{w^2}$.

\rahuld{The aforementioned discussion on the energy-transfer pathway, from TKE production of $\overline{u^2}$ to inter-component transfer to $\overline{w^2}$, explains the consistent variation exhibited by the outer region of $\overline{u^2}$ (figure \ref{fig2}b-d) and $\overline{uw}$ (figure \ref{fig2}i-l) with increasing $Re_{\tau}$. 
This consistency is noted for both ZPG and APG TBLs, suggesting that a comprehensive analysis of their $\overline{uw}$-characteristics could further unravel similarities between their energy transfer pathways.}
This would be particularly significant considering the correlation between $u$ and $w$ influences the magnitude and direction of the $\overline{u^2}$ wall-normal transport (${{\mathcal{T}}^{u}}$ = $-${${{\partial}{\overline{{u^2}w}}}/{{\partial}z}$}; figures \ref{fig2}w-y, \ref{fig3}j-l). 
To this end, the next section compares the wall-normal variation of $uw$-contributions from the dominant quadrants $\rm Q_2$ and $\rm Q_4$, with $\overline{u^2}$ and $\overline{{u^2}w}$. 
Notably, while both these quadrants contribute positively and significantly to ${{\mathcal{P}}^{u}}$, their contributions are significant but opposing for $\overline{{u^2}w}$ \citep{nagano1990}.

\section{Classification of the streamwise Reynolds-stress profiles based on energy-transfer mechanisms}
\label{results2}

\subsection{Empirical observations based on mean statistics}
\label{results2a}

\begin{figure}
   \captionsetup{width=1.0\linewidth}
\centering
\includegraphics[width=1.0\textwidth]{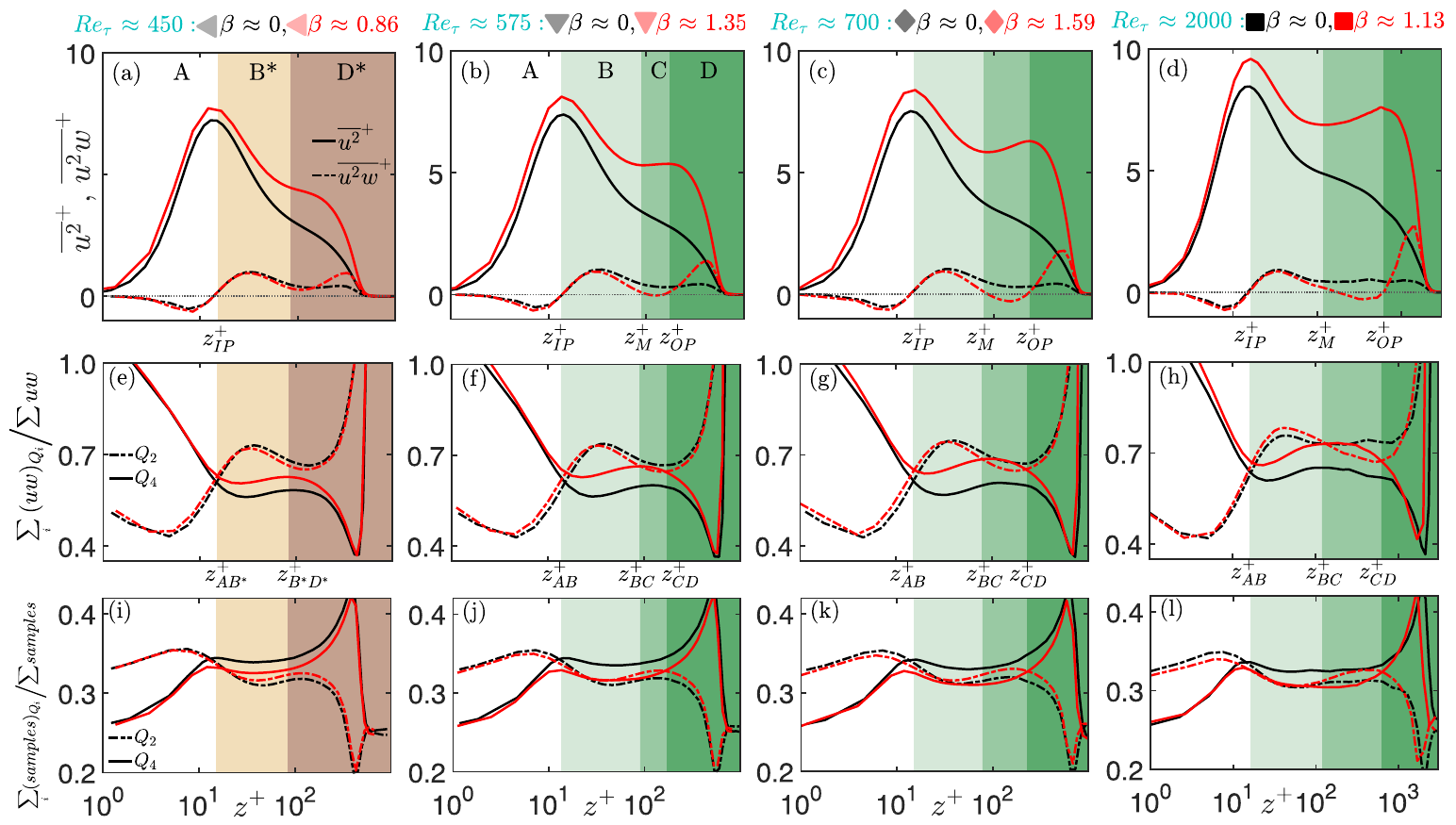}
\caption{(a-d) Streamwise Reynolds stress (${\overline{u^2}}^{+}$) and wall-normal flux of streamwise TKE (${\overline{{u^2}w}}^{+}$), (e-h) fractional contribution to Reynolds shear stress (${\overline{uw}}^{+}$) from quadrants $\rm Q_2$ and $\rm Q_4$, and (i-l) fraction of the total samples associated with $\rm Q_2$ and $\rm Q_4$, across various ZPG (black) and APG (red) TBL cases at matched $Re_{\tau}$ (see top of each column).
The background shading represents different regimes (A--D,${\textrm{B}}^*$,${\textrm{D}}^*$) defined in table \ref{tab2}, based on the relative predominance of $\rm Q_2$ or $\rm Q_4$ contributions to $\overline{uw}$ for APG TBLs.
Note that $z^+_{\textrm{AB}}$, $z^+_{{\textrm{A}}{B^*}}$, $z^+_{\textrm{BC}}$, $z^+_{\textrm{CD}}$ and $z^+_{{B^*}{D^*}}$ represent locations associated with regime transitions, while $z^+_{\textrm{IP}}$, $z^+_{\textrm{OP}}$ and $z^+_{M}$ represent the location of inner peak, outer peak and local minimum of the ${\overline{u^2}}^{+}$ profiles, respectively.}
\label{fig5}
\end{figure}

Figures \ref{fig5}(a-d) compare the wall-normal profiles of $\overline{u^2}$, and the wall-normal flux of streamwise TKE ($\overline{{u^2}w}$), for selected ZPG (in black) and APG TBL (in red) cases at matched $Re_{\tau}$.
Here, $z^+_{\textrm{IP}}$ and $z^+_{\textrm{OP}}$ are estimated following the same methodology as used previously for figure \ref{fig2}.
These profiles are compared with the fractional contributions from $\rm Q_2$ and $\rm Q_4$ quadrants to $\overline{uw}$ across the respective TBLs (figures \ref{fig5}e-h).
Their contributions are defined by following \citet{wallace2016} and computing ${\sum_{i}^{}}\:{(uw)_{Q_{i}}}$/${\sum_{}^{}}\:{uw}$ at each $z$, where the numerator represents summation across samples associated exclusively with $\rm Q_2$ ($u$ $<$ 0, $w$ $>$ 0) or $\rm Q_4$ ($u$ $>$ 0, $w$ $<$ 0) events.
Also plotted in figures \ref{fig5}(i-l) are the wall-normal variations for the fraction of the total samples associated with $\rm Q_2$ and $\rm Q_4$ events, in both the TBLs.
Combined together, all these figures depict a consistent connection between the behaviour of $\overline{u^2}$ and $\overline{{u^2}w}$ profiles, with statistical dominance of $\rm Q_2$ or $\rm Q_4$, which is discussed here.

\begin{figure}
   \captionsetup{width=1.0\linewidth}
\centering
\includegraphics[width=1.0\textwidth]{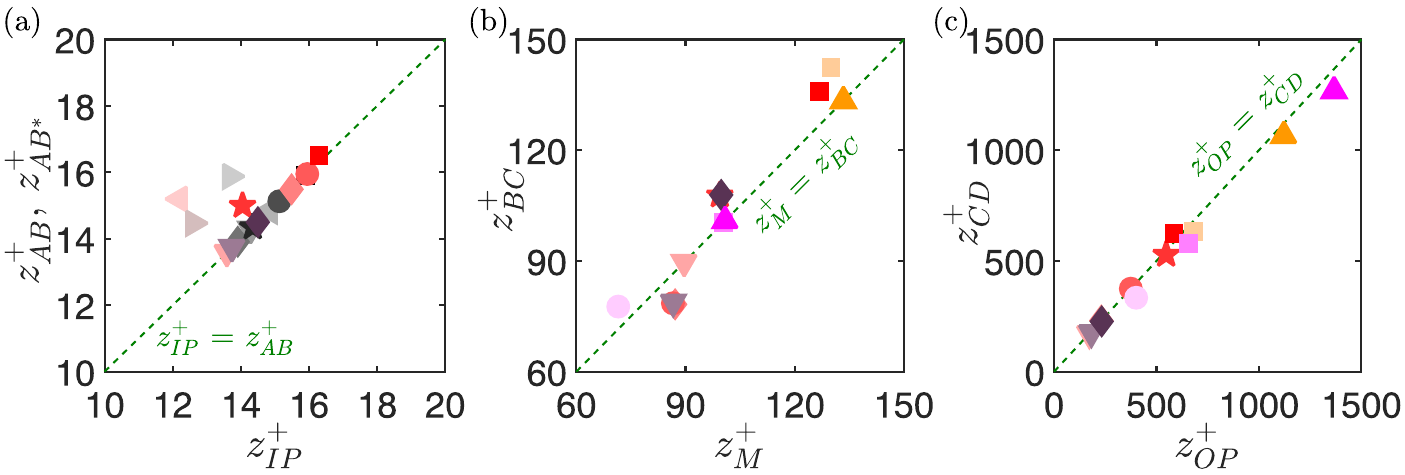}
\caption{Scatter plot demonstrating the one-to-one correlation between (a) $z^+_{\textrm{IP}}$ and $z^+_{\textrm{AB}}$ (or $z^+_{{\textrm{A}}{B^*}}$), (b) $z^+_{M}$ and $z^+_{\textrm{BC}}$, and (c) $z^+_{\textrm{OP}}$ and $z^+_{\textrm{CD}}$, associated with ${\overline{u^2}}^+$ and ${\overline{uw}}^+$ from various ZPG and APG TBL data sets. 
Definitions are given in $\S$\ref{results2} and figure \ref{fig5}. 
Symbols and color shading correspond to the various cases and data sets defined in figure \ref{fig1} and table \ref{tab1}.}
\label{fig6}
\end{figure}

Figures \ref{fig5}(a,e,i) compare statistics of a weak APG TBL with ZPG TBL at very low $Re_{\tau}$, and their respective wall-normal profiles look qualitatively similar.
For $z^+$ $<$ $z^+_{\textrm{IP}}$, ${\textrm{d}}{{\overline{u^2}}^{+}}$/${\textrm{d}}{z^+}$ $>$ 0, and this is associated with significantly higher contributions from $\rm Q_4$ events to $\overline{uw}$ than $\rm Q_2$ (figure \ref{fig5}e).
Interestingly, these statistically significant $\rm Q_4$ events correspond to a smaller fraction of the total $uw$-signal than $\rm Q_2$ (figure \ref{fig5}i), suggesting occurrence of intense `splatting' events in this near-wall region \citep{kim1987}.
This also results in the mean wall-normal flux of $\overline{u^2}$ directed towards the wall (\emph{i.e.}, $\overline{{u^2}w}$ $<$ 0).
The relative dominance of $\rm Q_4$ over $\rm Q_2$ reverses for $z^+$ $\gtrsim$ $z^+_{\textrm{IP}}$, which sees $\rm Q_2$ events contributing significantly more to $\overline{uw}$ and occurring across a smaller fraction of the total $uw$-signal (than $\rm Q_4$).
As a consequence, the mean wall-normal flux of $\overline{u^2}$ is directed away from the wall (\emph{i.e.}, $\overline{{u^2}w}$ $>$ 0), and interestingly, this region is associated with ${\textrm{d}}{{\overline{u^2}}^{+}}$/${\textrm{d}}{z^+}$ $<$ 0.
Based on the variation in predominance of $\rm Q_2$ and $\rm Q_4$, we can classify the near-wall region of ZPG TBLs and weak APG TBLs into regimes A (0 $\lesssim$ $z^+$ $\lesssim$ $z^+_{{\textrm{A}}{B^*}}$) and $B^*$ ($z^+_{{\textrm{A}}{B^*}}$ $\lesssim$ $z^+$ $\lesssim$ $z^+_{{B^*}{D^*}}$).
Here, $z^+_{{\textrm{A}}{B^*}}$ is the location for the `switch' in relative predominance between $\rm Q_2$ and $\rm Q_4$, and this is well known to coincide with $z^+_{\textrm{IP}}$ \citep{wallace2016,kim1987}.
Figure \ref{fig6}(a) reaffirms this one-to-one correspondence by comparing $z^+_{{\textrm{A}}{B^*}}$ and $z^+_{\textrm{IP}}$ for the various ZPG and weak APG TBL cases considered.
Further, as can be confirmed from the various $Re_{\tau}$ cases in figure \ref{fig5}, the above discussion on the behaviour of the flow is valid for all ZPG TBL cases ($Re_{\tau}$ $\lesssim$ 2000).
A noticeable difference between ZPG and weak APG TBLs, however, is that $\rm Q_4$ contributions to $\overline{uw}$ are significantly greater in the outer region of APG TBLs (figure \ref{fig5}e).
This is expected considering the previous findings of \citet{skaare1994turbulent} and \citet{lee2017}.
Next, we investigate how the dominance of $\rm Q_4$ varies with further energization of $\overline{u^2}$ in the APG TBL outer region.
To this end, we define $z^+_{{B^*}{D^*}}$ as the location corresponding to a local maximum of $\rm Q_4$ contributions in the outer region of ZPG and weak APG TBLs, and the regime $D^*$ defined across $z^+_{{B^*}{D^*}}$ $\lesssim$ $z^+$ $\lesssim$ ${\delta}^+_{99}$.

On considering an APG TBL at a moderately strong $\beta$ ($\approx$ 1.35) and slightly higher $Re_{\tau}$ in figures \ref{fig5}(b,f,j), an outer peak can be noted in the $\overline{u^2}$ profile.
Although the appearance of this outer peak is subtle, it is accompanied by distinct changes in the $\overline{{u^2}w}$ profile as well as enhancement of $\rm Q_4$ contributions to $\overline{uw}$.
Notably, there is a small intermediate regime at $z^+$ $\gg$ $z^+_{\textrm{IP}}$, wherein $\rm Q_4$ contributions to $\overline{uw}$ become greater than those from $\rm Q_2$ (figure \ref{fig5}f).
This enhanced $\rm Q_4$ strength is accompanied by a drop in the fraction of the $uw$-signal associated with $\rm Q_4$ events (figure \ref{fig5}j), suggesting occurrence of intense $\rm Q_4$ activity in the outer region of APG TBLs, analogous to the behaviour observed for $z^+$ $\lesssim$ $z^+_{\textrm{IP}}$.
This explains the mean wall-normal flux of $\overline{u^2}$ directed towards the wall in this intermediate regime (figure \ref{fig5}b), as also observed previously by \citet{skaare1994turbulent}.
On the other hand, the energy-transfer mechanisms in the inner region of this moderately strong APG TBL are similar to those noted in weak APG and ZPG TBLs.

\begin{table}
   \captionsetup{width=1.0\linewidth}
   \centering
  \begin{center} 
    \begin{tabular}{l|lllll}
      {\rotatebox[origin=c]{90}{Regimes}} & TBL & ${\rm Q_{i}}$ dominance & $z^+$-range & lower bound & upper bound \\    
      \noalign{\smallskip}\hline\noalign{\smallskip}

      \multirow{2}{*}{A} &
ZPG, weak APG & ${\overline{uw}}|_{\rm Q4}$ $>$ ${\overline{uw}}|_{\rm Q2}$ & 0 $\le$ $z^+$ $\le$ $z^+_{A{B^*}}$ & wall &
${\overline{uw}}|_{\rm Q4}$ $=$ ${\overline{uw}}|_{\rm Q2}$ \\
& moderate APG & ${\overline{uw}}|_{\rm Q4}$ $>$ ${\overline{uw}}|_{\rm Q2}$ & 0 $\le$ $z^+$ $\le$ $z^+_{AB}$ & wall &
${\overline{uw}}|_{\rm Q4}$ $=$ ${\overline{uw}}|_{\rm Q2}$ \\\hline
      
      $\rm B^*$ & ZPG, weak APG & ${\overline{uw}}|_{\rm Q2}$ $>$ ${\overline{uw}}|_{\rm Q4}$ & $z^+_{A{B^*}}$ $\le$ $z^+$ $\le$ $z^+_{{B^*}{D^*}}$ & ${\overline{uw}}|_{\rm Q4}$ $=$ ${\overline{uw}}|_{\rm Q2}$ & maximum ${\overline{uw}}|_{\rm Q4}$
 \\\hline

      B & moderate APG & ${\overline{uw}}|_{\rm Q2}$ $>$ ${\overline{uw}}|_{\rm Q4}$ & $z^+_{AB}$ $\le$ $z^+$ $\le$ $z^+_{BC}$ &
${\overline{uw}}|_{\rm Q4}$ $=$ ${\overline{uw}}|_{\rm Q2}$ & ${\overline{uw}}|_{\rm Q4}$ $=$ ${\overline{uw}}|_{\rm Q2}$ \\
      \hline
      C & moderate APG & ${\overline{uw}}|_{\rm Q4}$ $>$ ${\overline{uw}}|_{\rm Q2}$ & $z^+_{BC}$ $\le$ $z^+$ $\le$ $z^+_{CD}$ &
${\overline{uw}}|_{\rm Q4}$ $=$ ${\overline{uw}}|_{\rm Q2}$ & ${\overline{uw}}|_{\rm Q4}$ $=$ ${\overline{uw}}|_{\rm Q2}$ \\
      \hline
      D & moderate APG & ${\overline{uw}}|_{\rm Q2}$ $>$ ${\overline{uw}}|_{\rm Q4}$ & $z^+_{CD}$ $\le$ $z^+$ $\le$ ${\delta}^+$ &
${\overline{uw}}|_{\rm Q4}$ $=$ ${\overline{uw}}|_{\rm Q2}$ & TBL edge \\ 
      \hline
      ${\rm D^*}$ & ZPG, weak APG & ${\overline{uw}}|_{\rm Q2}$ $>$ ${\overline{uw}}|_{\rm Q4}$ & $z^+_{{B^*}{D^*}}$ $\le$ $z^+$ $\le$ ${\delta}^+$ &
maximum ${\overline{uw}}|_{\rm Q4}$ & TBL edge \\
      \hline
    \end{tabular}
  \end{center}
    \caption{Table summarizing the definitions of various regimes classified in $\S$\ref{results2a}, based on the relative dominance of $\rm Q_2$ or $\rm Q_4$ Reynolds shear stress events in both ZPG and APG TBLs.}   
  \label{tab2}
\end{table}

The dominance of $\rm Q_4$ over $\rm Q_2$ in an intermediate regime, in addition to the very near-wall region, means that the $\overline{u^2}$ profiles of moderately strong APG TBLs can be classified into four regimes. 
The regions associated with predominance of $\rm Q_4$ are defined as regimes A (0 $\lesssim$ $z^+$ $\lesssim$ $z^+_{\textrm{AB}}$) and C ($z^+_{\textrm{BC}}$ $\lesssim$ $z^+$ $\lesssim$ $z^+_{\textrm{CD}}$), while those corresponding to predominance of $\rm Q_2$ are defined as regimes B ($z^+_{\textrm{AB}}$ $\lesssim$ $z^+$ $\lesssim$ $z^+_{\textrm{BC}}$) and D ($z^+_{\textrm{CD}}$ $\lesssim$ $z^+$ $\lesssim$ ${\delta}^+_{99}$).
Table \ref{tab2} summarizes the definitions of these limits, which are used to classify the various regimes based on the relative dominance of $\rm Q_2$ or $\rm Q_4$.
Here, $z^+_{\textrm{BC}}$ and $z^+_{\textrm{CD}}$ correspond to the two locations in the outer region at which there is a switch in the relative predominance of $\rm Q_4$ and $\rm Q_2$ contributions, while $z^+_{\textrm{AB}}$ corresponds to the same in the inner region (figure \ref{fig5}f).
Analogous to the correlations noted in the inner region, where $z^+_{\textrm{IP}}$ coincides with $z^+_{\textrm{AB}}$, we find $z^+_{\textrm{CD}}$ to coincide with $z^+_{\textrm{OP}}$ in the outer region.
In between the two peaks, there is a minimum in the $\overline{u^2}$ profile at $z^+_M$ (= ${{\textrm{d}}^2}{{\overline{u^2}}^{+}}$/${\textrm{d}}{{z^+}^2}$ $>$ 0 and ${\rm d}{{\overline{u^2}}^{+}}$/${\rm d}{z^+}$ = 0), which is found to coincide with $z^+_{\textrm{BC}}$.
Remarkably, both regimes A and C are associated with $\overline{{u^2}w}$ $<$ 0 and ${\textrm{d}}{{\overline{u^2}}^{+}}$/${\textrm{d}}{z^+}$ $>$ 0, while regimes B and D are associated with $\overline{{u^2}w}$ $>$ 0 and ${\textrm{d}}{{\overline{u^2}}^{+}}$/${\textrm{d}}{z^+}$ $<$ 0.
\rahuld{Hence, it can be argued that the wall-normal energy transport across the `outer' region ($z^+_{\textrm{BC}}$ $\lesssim$ $z^+$ $\lesssim$ ${\delta}^+_{99}$) is phenomenologically similar to that across the `inner' region (0 $\lesssim$ $z^+$ $\lesssim$ $z^+_{\textrm{BC}}$), with $\rm Q_4$ and $\rm Q_2$ respectively dominating below and above each of the $\overline{u^2}$ peaks.
This is consistent with previous discussions in $\S$\ref{results1}, based on analysis of the Reynolds-stress-transport terms, wherein the energy-transfer pathways were found to be similar near both the inner and outer peaks of $\overline{u^2}$.}
Here, note that the use of `inner' and `outer' is not per the classical definitions in the literature.
Interestingly, this connection between the sign of ${\textrm{d}}{{\overline{u^2}}^{+}}$/${\textrm{d}}{z^+}$, and the relative dominance of $\rm Q_4$ or $\rm Q_2$, has also been noted previously in rough wall flows \citep{katul2006}, hinting at universality in the influence of these quadrants on $\overline{u^2}$.

The aforementioned hypothesis can be tested by analyzing the same statistics, at similar moderately strong APGs ($\beta$ $\approx$ 1.59, 1.13), in TBLs at relatively high $Re_{\tau}$ values of 700 (figures \ref{fig5}c,g,k) and 2000 (figures \ref{fig5}d,h,l).
A higher $Re_{\tau}$ leads to increased separation between scales associated with the inner and outer peaks, which should result in widening of the intermediate regime C ($z^+_{\textrm{BC}}$ $\lesssim$ $z^+$ $\lesssim$ $z^+_{\textrm{CD}}$).
As a consequence, $z^+_{\textrm{IP}}$, $z^+_{\textrm{OP}}$ as well as $z^+_{\rm M}$ can be distinctly noted in each of these relatively high $Re_{\tau}$ cases, and indeed these maxima and minima locations respectively coincide with ($z^+_{\textrm{AB}}$, $z^+_{\textrm{CD}}$) and $z^+_{\textrm{BC}}$ defined based on relative predominance of $\rm Q_2$ or $\rm Q_4$ (table \ref{tab2}).
This is reaffirmed in figures \ref{fig6}(a-c), which demonstrate consistency in $\overline{u^2}$ profile classification across all moderately strong APG TBL cases, despite their different upstream histories.
Figures \ref{fig5}(c,g,k) and \ref{fig5}(d,h,l), thus, confirm that the underlying wall-normal energy-transport mechanisms discussed for figures \ref{fig5}(b,f,j) remain valid for all moderately strong APG scenarios.
That is, regimes A and C correspond to dominant contributions of $\rm Q_4$ to $\overline{uw}$, which are associated with smaller fractions of the total $uw$-signal (compared to $\rm Q_2$).
The mean wall-normal flux of $\overline{u^2}$, consequently, is directed towards the wall in both these regimes, and these again coincide with ${\textrm{d}}{{\overline{u^2}}^{+}}$/${\textrm{d}}{z^+}$ $>$ 0.
On the other hand, the flow phenomena is entirely opposite in regimes B and D, where the $\rm Q_2$ events predominate over $\rm Q_4$ events leading to $\overline{{u^2}w}$ $>$ 0. 
Hence, phenomenologically, $z^+_{\textrm{BC}}$ appears to be the intermediate location where wall-ward $\overline{u^2}$ flux from the outer, and away-from-wall flux from the inner region `equalize', plausibly yielding a minimum in $\overline{u^2}$.

The quadrant analysis in this section reveals that the mean wall-normal flux of $\overline{u^2}$ is characteristically different in the outer region of moderately strong APG TBLs, compared to that in low-$Re_{\tau}$ ZPG and weak-APG TBLs.
However, this contrasts with the similar energy transfer pathways noted for all these TBLs in $\S$\ref{results1}.
We hypothesize this inconsistency is owing to the statistically weak TKE production in the outer region of low $Re_{\tau}$ ZPG and weak-APG TBLs, owing to which the associated energy transfer mechanisms get masked by much stronger transfer mechanisms originating in their inner region (owing to stronger TKE production).
We investigate this hypothesis next via conditional analysis, to check if the dominance of $\rm Q_4$ events noted in moderately strong APG TBLs, is also unravelled on conditionally averaging the outer regions of ZPG and weak-APG TBLs.

\subsection{Conditional analysis of the outer region}
\label{results2b}

Here, we report conditional averaging based on statistically significant TKE production events detected in the outer region, to bring out the energy-transfer processes associated with the outer region.
This is achieved by deploying the same time-series thresholding criteria as that used by \citet{deshpande2021uw} and \citet{narasimha2007}, \emph{i.e.} $|{uw}(z;t)|$ $>$ ${k}(uw(z;t))_{\rm sd}$, which identifies time-series segments associated with strong Reynolds shear-stress events (\emph{i.e.}, contributing to strong TKE production).
Here, $(uw(z;t))_{\rm sd}$ represents the standard deviation of the $uw$-signal at $z$, $t$ represents time and $k$ represents the constant threshold, following the original proposal of \citet{narasimha2007}.
Since we are interested in analyzing the wall-normal profiles of conditional-averaged statistics, the entire flow field will be conditioned based on the $uw$-signal at a single $z$-location.
For the present analysis, $k$ = 0.50 is chosen, however we have checked that the conclusions do not depend on the threshold when $k$ $\gtrsim$ 0.25.
The threshold, $k$ = 0.25, was found by \citet{deshpande2021uw} to recover 99\% of the mean Reynolds shear stress despite neglecting $\sim$43\% of the total $uw$-signal (meaning this 43\% essentially contributed insignificantly to $\overline{uw}$).
Here, since we want to extract only the most statistically dominant $uw$-signal, we have opted for $k$ = 0.50 that leads to dis-consideration of $\sim$62\% of the $uw$-signal.

\begin{figure}
   \captionsetup{width=1.0\linewidth}
\centering
\includegraphics[width=1.0\textwidth]{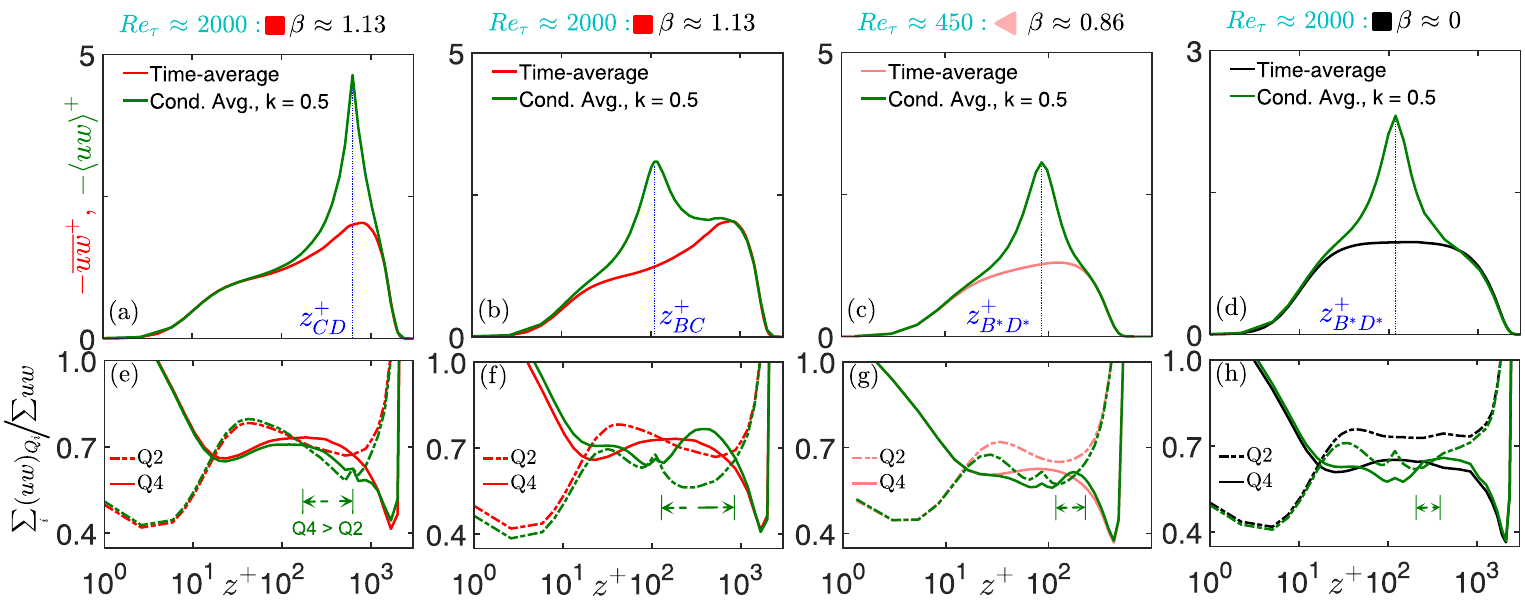}
\caption{Time- (in red or black)  and conditionally-averaged (in green) (a-d) Reynolds shear stresses and (e-h) fractional contributions to Reynolds shear stresses from $\rm Q_2$ and $\rm Q_4$, plotted for various ZPG and APG TBL cases considered previously in figure \ref{fig5}.
The flow-conditioning criteria has been adopted from \citet{deshpande2021uw} and defined in $\S$\ref{results2b}, with the $z-$location for conditioning indicated by the vertical blue dashed lines in (a-d).
Dashed green lines with arrows in (e-h) indicate $z^+$-range where $\rm Q_4$ $>$ $\rm Q_2$ contributions to ${\langle}uw{\rangle}$.}
\label{fig7}
\end{figure}

We begin by first demonstrating the utility of this conditional-averaging analysis by applying it to a moderately strong APG TBL case at $Re_{\tau}$ $\approx$ 2000, for which similarity in $\overline{u^2}$-wall-normal transport between the inner and outer regions has already been demonstrated (illustrated by the relative dominance of $\rm Q_2$ and $\rm Q_4$ in figure \ref{fig5}d).
Figures \ref{fig7}(a,e) report the conditionally averaged statistics (indicated by angular brackets; plotted using green lines) that are compared with the time-averaged Reynolds shear stresses as well as fractional contributions from $\rm Q_4$ and $\rm Q_2$ to $\overline{uw}$ (same as those plotted in figures \ref{fig5}d,h).
For figures \ref{fig7}(a,e), the $z$-location for conditioning is chosen same as the location of the outer peak, \emph{i.e.} $z^+_{\rm CD}$ = $z^+_{\rm OP}$, and we can note the conditionally-averaged $\rm Q_4$ and $\rm Q_2$ contributions to the Reynolds shear stress to be qualitatively similar to that noted in the time-averaged statistics.
To demonstrate the choice of flow conditioning location does not influence this analysis, it is repeated on the same data set but conditioned at the local maxima of $\rm Q_4$ contributions (figure \ref{fig7}b,f), which is close to $z^+_{\rm BC}$ for moderately strong APG TBLs.
While the conditionally-averaged statistics change quantitatively on changing the conditioning location from $z^+_{\rm CD}$ to $z^+_{\rm BC}$, the relative dominance of the $\rm Q_4$ and $\rm Q_2$ contributions remains the same qualitatively, and also in the same wall-normal range as reported based on the time-averaged statistics.
For example, the dominance of $\rm Q_4$ over $\rm Q_2$ in the outer region is found to exist in a similar $z^+$-range in both figures \ref{fig7}(e,f), which is indicated by the dashed green lines with arrows.
Importantly, independence from the specific choice of flow conditioning location permits us to deploy this analysis technique next on to ZPG and weak-APG TBL data, \emph{i.e.} without discernible $\overline{u^2}$ outer peaks or dominance of $\rm Q_4$ in the outer region.

Figures \ref{fig7}(c,g) depict the conditionally-averaged statistics computed for a weak-APG TBL case shown previously in figures \ref{fig5}(a,e,i), for which the flow is conditioned at the local maxima of $\rm Q_4$ contributions (\emph{i.e.}, $z^+_{{B^*}{D^*}}$).
Conditioning the flow based on statistically significant $uw$-signals, in the outer region, indeed brings out the dominance of $\rm Q_4$ over $\rm Q_2$ contributions to ${\langle}uw{\rangle}$ (marked by green arrows), which is otherwise not observed in the time-averaged statistics.
The same analysis is then extended onto ZPG TBL data at $Re_{\tau}$ $\approx$ 2000 (figures \ref{fig7}d,h).
Note that the time-averaged statistics for this ZPG case were previously plotted in figures \ref{fig5}(d,h).
Remarkably, \rahuld{conditionally averaging the flow at $z^+_{{B^*}{D^*}}$ again reveals the dominance of $\rm Q_4$ over $\rm Q_2$ contributions to ${\langle}uw{\rangle}$, providing evidence in support of this hypothesis from \citet{mklee2019} and \citet{deshpande2021uw}. 
This unravels the existence of similar wall-normal energy-transfer mechanisms in the outer region, as those noted in the inner region, for both ZPG and APG TBLs.}
It also confirms that energy-transfer mechanisms in the outer region of low-$Re_{\tau}$ ZPG and weak-APG TBLs are statistically weak in a time-averaged sense, and are consequently masked by the much stronger mechanisms associated with the inner region.

\section{Conceptual picture of the streamwise energy-transfer mechanisms}
\label{results2c}

\begin{figure}
   \captionsetup{width=1.0\linewidth}
\centering
\includegraphics[width=0.95\textwidth]{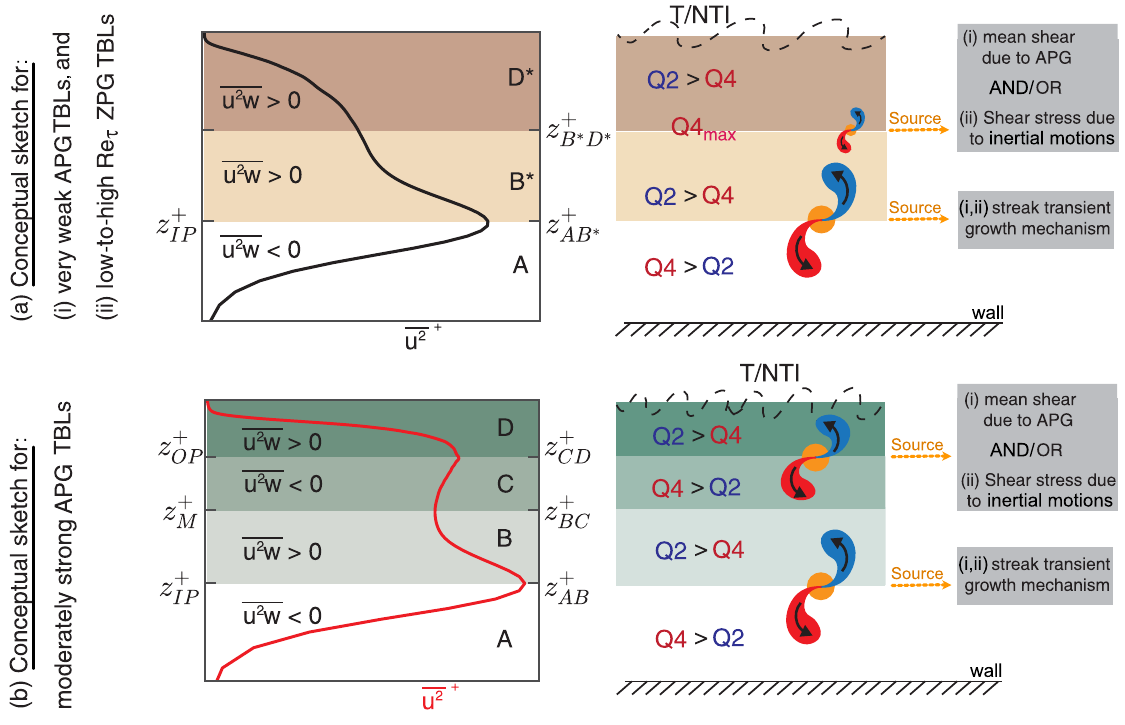}
\caption{Conceptual sketches describing the energy-transfer mechanisms inferred based on figures \ref{fig2}-\ref{fig7}, for (a) weak APG and low-to-high $Re_{\tau}$ ZPG TBLs, and (b) moderately strong APG TBLs. 
Solid golden circles indicate $z^+$ locations where $\rm Q_2$ and $\rm Q_4$ events are equally strong and occupy the same fraction of the total signal. 
These solid circles are referred as `epicentres' owing to the maxima in ${\mathcal{P}}^u$ and $-{{\Pi}^u}$.
T/NTI refers to the turbulent/non-turbulent interface.} 
\label{fig8}
\end{figure}

Having gained significant insights based on the empirical trends reported in figures \ref{fig2}-\ref{fig7}, here we attempt to represent the relationship between key energy-transfer mechanisms and $\overline{u^2}$ profiles via a generalized conceptual sketch.
This will be discussed first for APG TBLs, and then for ZPG TBLs.
Developing this general understanding will help to propose a hypothesis for energy-transfer mechanisms in very high $Re_{\tau}$ ZPG TBLs, \rahul{for future investigations}.
In figure \ref{fig8}(b), the energy-transfer mechanisms in moderately strong APG TBLs are depicted to be governed by two dominant `epicentres', at $z^+_{\textrm{AB}}$ (= $z^+_{\textrm{IP}}$) and $z^+_{\textrm{CD}}$ (= $z^+_{\textrm{OP}}$).
These epicentres correspond to the coexisting local maxima noted for ${\mathcal{P}}^u$ and $-{{\Pi}^u}$, as well as equal contributions from $\rm Q_4$ and $\rm Q_2$ to $\overline{uw}$, in both the inner and outer regions ($\S$\ref{results1} and $\S$\ref{results2}).
This justifies referencing them as epicentres, considering these locations correspond to intense $\overline{u^2}$ production as well as its transfer to $\overline{w^2}$ (via ${\Pi}^u$) and wall-normal transport ($\overline{{u^2}w}$).
However, by no means these should be interpreted as the only $z^+$-locations for TKE production and inter-component energy transfer, which is well-known to occur across a range of scales and wall-normal locations \citep{cossu2017,mklee2019}.
In figure \ref{fig8}(b), $z^+_{\textrm{AB}}$ is the epicentre in the near-wall region, associated with the energy of the inner-scaled streaks (which is part of the near-wall cycle proposed by \citealp{hamilton1995}).
While, $z^+_{\textrm{CD}}$ is the epicentre in the outer region and is associated with either the mean shear or shear stresses based on inertial motions (depending on the $Re_{\tau}$).
The concentration of intense streamwise TKE production at both these locations explains their coincidence with $\overline{u^2}$ peaks, and this results in the positive and negative gradients of $\overline{u^2}$ below and above either of these peaks, respectively.
The TKE produced near these epicentres is respectively transported towards and away from the wall by $\rm Q_4$ ($w$ $<$ 0) and $\rm Q_2$ ($w$ $>$ 0) events.
This explains the relative predominance of $\rm Q_4$ and $\rm Q_2$ below and above these TKE production peaks, respectively.
The wall-ward flux of $\overline{u^2}$ from the outer epicentre (carried by $\rm Q_4$), merges with the flux directed away from the wall ($\rm Q_2$), from the inner epicentre, at an intermediate location $z^+_{\textrm{BC}}$ (= $z^+_{M}$).
This explains the coincidence of the $\overline{u^2}$ minima with the balance between $\rm Q_2$ and $\rm Q_4$ contributions at $z^+_{\textrm{BC}}$, distinguishing it from the epicentres at $z^+_{\textrm{AB}}$ and $z^+_{\textrm{CD}}$.
In this way, figure \ref{fig8}(b) summarizes the classification of the $\overline{u^2}$ profiles (into regimes A--D) based on the energy-transfer mechanisms noted for moderately strong APG TBLs, found to be independent of the history effects.

On the other hand, a weak APG implies that the TKE production in the outer region is not `intense enough' to generate an outer peak in the $\overline{u^2}$ profile (figure \ref{fig2}), a hypothesis that was validated in $\S$\ref{results2b}.
We found that although there is a weak epicentre in the outer region (figure \ref{fig8}a) associated with significant TKE production (figure \ref{fig2}f), the wall-ward flux of $\overline{u^2}$ from this epicentre (via $\rm Q_4$ events) is not strong enough to balance against the $\overline{u^2}$ flux away from the wall, carried by $\rm Q_2$ events originating near $z^+_{{\textrm{A}}{B^*}}$.
As a consequence, $\rm Q_2$ $>$ $\rm Q_4$ for all $z^+$ $>$ $z^+_{{\textrm{A}}{B^*}}$ in weak APG TBLs.
Hence, based on the evidences from time-averaged (figures \ref{fig2}-\ref{fig6}) and conditionally-averaged statistics (figure \ref{fig7}) of weak and moderately strong APG TBLs, we argue that the presence of a $\overline{u^2}$ outer peak directly corresponds with the predominance of $\rm Q_4$ contributions to $\overline{uw}$ in an intermediate regime C.
Indeed, the emergence of the $\overline{u^2}$ outer peak with an increase in APG strength can be tracked by investigating the growth of $\rm Q_4$ contributions at $z^+_{{B^*}{D^*}}$ (\emph{i.e.} maxima of $\rm Q_4$), which increases beyond the contributions from $\rm Q_2$ as the outer peak emerges (figure \ref{fig5}a-h).

Next, we extend the same argument to ZPG TBLs, to discuss the \rahul{extent of similarity in energy-transfer mechanisms with APG TBLs.}
While the mean shear in the outer region of ZPG TBLs is not as strong as in the case of APGs, an increase in $Re_{\tau}$ leads to the energization and broadening of the inertial eddy hierarchy, which increases the TKE production in the outer region via growth in the Reynolds shear stresses ($\S$\ref{results1}).
\rahuld{Based on figures \ref{fig2}-\ref{fig5}, it is reasonable to conclude that the energy-transfer mechanisms in a low-to-high $Re_{\tau}$ ZPG TBL (where no distinct $\overline{u^2}$ outer peak is observed) are analogous to those in a weak APG TBL, which has been conceptualized in figure \ref{fig8}(a).}
As demonstrated via figures \ref{fig7}(d,h) for a moderate $Re_{\tau}$ $\sim$ 2000 of a ZPG TBL, the streamwise TKE production in the outer region is statistically insignificant at this $Re_{\tau}$, owing to which the wall-ward flux of $\overline{u^2}$ energy (from the outer epicentre) is not strong enough to yield predominance of $\rm Q_4$ contributions in the intermediate region (over $\rm Q_2$ originating from the inner region).
However, it is plausible that significant increase in $Re_{\tau}$ could enhance the outer TKE production \citep{marusic2010high} such that the energy-transfer mechanisms for very high $Re_{\tau}$ ZPG TBLs become analogous to those conceptualized in figure \ref{fig8}(b), for moderately strong APG TBLs.
\rahul{This hypothesis, however, can only be confirmed in the future after availability of canonical data sets at very high $Re_{\tau}$.}

\section{Summary and concluding remarks}
\label{conclude}

This study investigates energy-transfer mechanisms associated with the $Re_{\tau}$-increment of the streamwise Reynolds stress profiles ($\overline{u^2}$) in the outer region of TBLs. 
Published data sets of weak and moderately strong APG TBLs are analyzed alongside those of ZPG TBLs, to offer phenomenological arguments based on the energy-transfer mechanisms prominent in the outer region.
Connections between the wall-normal profiles of $\overline{u^2}$, its production (${\mathcal{P}}^{u}$), inter-component (${\Pi}^u$) and wall-normal transport (${\mathcal{T}}^u$), as well as viscous dissipation (${\mathcal{E}}^u$) are investigated, all of which increase in statistical significance with increasing $Re_{\tau}$ \citep{marusic2010high,cho2018,mklee2019}.
\rahuld{The present analysis reveals similarity in the energy-transfer pathways between the inner and outer regions of ZPG and APG TBLs.
The energy pathway originates from the production of $\overline{u^2}$ via ${\mathcal{P}}^u$, which is then either (i) transferred to $\overline{v^2}$ or $\overline{w^2}$ by ${\Pi}^u$, (ii) transported in the wall-normal direction by ${\mathcal{T}}^u$ or (iii) lost to heat through ${\mathcal{E}}^u$.
The enhancement of these energy-transfer processes (with $Re_{\tau}$) for both ZPG and APG TBLs is predominantly associated with the energization and broadening of the inertial eddy hierarchy, which is indicated by the growth of the Reynolds shear stresses.}
This similarity noted between ZPG and low-to-moderately strong APG TBLs is consistent with previous findings of \citet{gungor2022}.
They obtained similar conclusions on comparing transfer mechanisms between canonical channel flows and APG TBLs, thereby supporting extrapolation of present findings from APG TBLs on to all canonical flows.

For moderately strong APGs, which are characterized by both inner and outer $\overline{u^2}$ peaks, similar energy-transfer mechanisms are found to be centred around either of these peaks (in a phenomenological sense).
Both inner ($z^+_{\textrm{AB}}$) and outer peaks ($z^+_{\textrm{CD}}$) correspond with a local maximum in ${\mathcal{P}}^u$ and ${\Pi}^u$ profiles, and are accompanied by the predominance of $\overline{u^2}$ energy flux ($\overline{{u^2}w}$) towards/away from the wall in the region immediately below/above each of them.
This inspires the classification of the $\overline{u^2}$ profile into four distinct regimes, the definition of which is independent of the upstream history effects.
The `inner' region has an epicentre of ${\mathcal{P}}^u$ and ${\Pi}^u$ at $z^+_{AB}$ that regulates $\overline{u^2}$ energy flux towards (in regime A; 0 $\lesssim$ $z^+$ $\lesssim$ $z^+_{AB}$) and away from the wall (in regime B; $z^+_{AB}$ $\lesssim$ $z^+$ $\lesssim$ $z^+_{BC}$).
This wall-normal transport is governed respectively by the $\rm Q_4$ and $\rm Q_2$ quadrants of the Reynolds shear stress.
A similar conceptual picture is observed in the `outer' region, with the energy-transfer mechanisms maximum around the outer epicentre $z^+_{CD}$, which defines the other two regimes: C ($z^+_{BC}$ $\lesssim$ $z^+$ $\lesssim$ $z^+_{CD}$) and D ($z^+_{CD}$ $\lesssim$ $z^+$ $\lesssim$ ${\delta}^+_{99}$).
These two `inner' and `outer' regions merge at an intermediate location corresponding to the minimum in the $\overline{u^2}$ profiles, which is located where the $\overline{u^2}$ energy flux from the two epicentres `equalize'.

In scenarios of low-to-high $Re_{\tau}$ ZPG TBLs and weak APG TBLs, however, no distinct outer peak is noted in $\overline{u^2}$, ${\mathcal{P}}^u$ or ${\Pi}^u$, with the wall-normal flux of $\overline{u^2}$ directed away from the wall (\emph{i.e.}, $\rm Q_2$ dominance) across the outer region.
\rahuld{Considering, however the similarity exhibited by the energy-transfer pathways between moderately strong APG TBLs and ZPG TBLs, it is hypothesized that the outer energy mechanisms in the latter are statistically weaker than those in their inner region.}
Conditional-averaging is implemented to unravel the flow phenomena associated exclusively with the outer region energy mechanisms, in weak APG and ZPG TBLs;
it reveals relative dominance of $\rm Q_4$ and $\rm Q_2$ in their outer region to be consistent with that observed for moderately strong APG TBLs.
\rahul{These trends suggest that a $\overline{u^2}$ outer peak can emerge for ZPG TBLs only at very high $Re_{\tau}$, when the outer-energy mechanisms become sufficiently significant to overcome the influence of the inner-energy mechanisms.
This hypothesis, however, can only be confirmed in the future after availability of well-resolved measurements at very high $Re_{\tau}$, potentially unravelling new flow physics associated with these energy-transfer mechanisms.}

To conclude, the present study improves our fundamental understanding of both ZPG and APG TBLs on several fronts.
Namely, it brings to light, a direct connection between the $z$-variations in their $\overline{u^2}$ profiles and the Reynolds shear stress-carrying ($\rm Q_4$,$\rm Q_2$) events, and the influential role played by the latter in the $Re_{\tau}$-variation of the former.
\rahuld{Further, this study also suggests that moderately strong APG TBLs can be potentially used to qualitatively understand the trends in energy-transfer mechanisms in very high $Re_{\tau}$ ZPG TBLs.}
Since the fundamental mechanisms causing the energization of the outer region are typically centred at distinctly different locations in these two TBL types, a combination of both parameters (\emph{i.e.}, high $Re_{\tau}$ APG TBLs) will be expected to have interesting flow physics \citep{vinuesa2017wing,romero2022properties,deshpande2023}.
The present analysis and arguments, hence, provide a plausible approach towards understanding this unexplored regime of weak-to-moderately strong APG TBLs at high $Re_{\tau}$.

\section*{Acknowledgments}
The authors are grateful to Drs. R. Pozuelo and C. Sanmiguel Vila for sharing their data sets, and thank Profs. H. Nagib, I. Marusic, A. Smits and M. Hultmark for their encouraging comments and helpful discussions.
R. Deshpande is grateful to the University of Melbourne's Postdoctoral Fellowship that funded his visit to KTH Stockholm.
R. Deshpande also acknowledges the Office of Naval Research (ONR) and ONR Global grant\# N62909-23-1-2068.
R. Vinuesa acknowledges the financial support from ERC grant\# `2021-CoG-101043998, DEEPCONTROL'. 
Views and opinions expressed are however those of the authors only and do not necessarily reflect those of the European Union or the European Research Council. 
Neither the European Union nor the granting authority can be held responsible for them.

\section*{Declaration of Interests} 

The authors report no conflict of interest.

\bibliographystyle{jfm}
\bibliography{ReynoldsShearStress_u2prof_bib}

\end{document}